\documentclass[10pt]{article}
\usepackage{graphicx}
\usepackage{amsmath}
\usepackage{amssymb}
\usepackage{caption2}
\begin{document}
\begin{center}
{\large {\bf \sc{ Analysis of the tetraquark and hexaquark molecular states with the QCD sum rules  }}} \\[2mm]
Zhi-Gang  Wang \footnote{E-mail: zgwang@aliyun.com.  }   \\
 Department of Physics, North China Electric Power University, Baoding 071003, P. R. China
\end{center}

\begin{abstract}
In this article,  we construct the color-singlet-color-singlet type currents and the  color-singlet-color-singlet-color-singlet type currents  to study the scalar $D^*\bar{D}^*$, $D^*D^*$ tetraquark molecular states and the vector  $D^*D^*\bar{D}^*$, $D^*D^*D^*$ hexaquark molecular states  with the QCD sum rules in details. In calculations, we choose the pertinent  energy scales of the QCD spectral densities  with the  energy scale formula $\mu=\sqrt{M^2_{T}-(2{\mathbb{M}}_c)^2}$  and $\sqrt{M^2_{H}-(3{\mathbb{M}}_c)^2}$ for the tetraquark and hexaquark molecular states respectively in a consistent way.
 We obtain stable QCD sum rules for the scalar  $D^*\bar{D}^*$, $D^*D^*$ tetraquark molecular states and the vector $D^*D^*\bar{D}^*$ hexaquark molecular state, but cannot obtain stable QCD sum rules for the vector $D^*D^*D^*$ hexaquark molecular state.
 The connected (nonfactorizable) Feynman diagrams at the tree level (or the lowest order) and their induced diagrams via substituting the quark lines  make positive contributions  for the scalar $D^*D^*$ tetraquark molecular state, but  make negative or destructive  contributions for the vector  $D^*D^*D^*$ hexaquark molecular state.  It is of no use or meaningless to distinguish the factorizable and nonfactorizable properties of the Feynman diagrams in the color space in the operator product expansion so as to interpret  them in terms of the hadronic observables,  we can only obtain information about the short-distance and long-distance contributions.

\end{abstract}



\section{Introduction}
The QCD sum rules approach, which developed about forty years ago by Shifman,  Vainshtein and  Zakharov, has become a widely used theoretical
 tool in studying the hadron properties, such as the masses, decay constants, form-factors, coupling constants, light-cone distribution amplitudes, etc
 \cite{SVZ79,Reinders85,Colangelo-Review}. We carry out the operator product expansion for the correlation functions $\Pi(p^2)$
  in the deep Euclidean space, $P^2=-p^2\gg \Lambda_{QCD}^2$, $t\sim \vec{x}\sim \frac{1}{\sqrt{P^2}}$, $x^2\sim \frac{1}{P^2}$, which represent
   the short-distance quark-antiquark fluctuations and can be treated in perturbative QCD.
  A certain energy scale $\mu$ is necessary to separate the regions of the short distance and long distance, the short-distance quark-gluon interactions at
 the momenta greater than  $\mu^2$ are included in the Wilson's coefficients, while the long-distance quark-gluon interactions or soft quark-gluon effects at
the momenta less than $\mu^2$ are absorbed into the vacuum condensates, which have universal values and are applicable in all the QCD sum rules.
    On the other hand, at the positive $p^2$, the correlation functions $\Pi(p^2)$ can be expressed in terms of hadronic observables.
    The  correlation functions $\Pi(p^2)$ obtained at an arbitrary point $p^2<0$ relate to the hadron representation through dispersion relation.

Experimentally, a number of charmonium-like and bottomonium-like states, in other words the exotic $X$, $Y$ and $Z$ states, were observed after the observation  of the $X(3872)$  by the  Belle collaboration \cite{X3872-2003}.  In 2006, R. D. Matheus et al assigned  the $X(3872) $ to be  the  diquark-antidiquark type  tetraquark state with the spin-parity-charge-conjugation $J^{PC}=1^{++}$, and studied  its mass with the QCD sum rules \cite{Narison-3872}.  Thereafter  the QCD sum rules become a powerful theoretical approach in studying the exotic $X$, $Y$, $Z$ and $P_c$ states and have  given many successful descriptions of the hadron properties, such as the masses and decay widths \cite{MNielsen-review-1812}. As the exotic $X$, $Y$, $Z$ and $P_c$ states lie near the meson-meson or meson-baryon thresholds, we can  assign them to be the tetraquark or pentaquark molecular states naively and intuitively, and interpolate them  with the color-singlet-color-singlet type currents  \cite{QCDSR-tetra-mole-mass,WangZG-4-quark-mole,QCDSR-tetra-mole-width,QCDSR-Pc-mole}.

The color-singlet-color-singlet type currents, which consist of four valance quarks, couple potentially to the tetraquark molecular states, however, the quantum field theory does not forbid  the non-vanishing couplings between the  color-singlet-color-singlet type currents and meson-meson scattering states.
For the color-singlet-color-singlet type currents,   Lucha, Melikhov and Sazdjian assert that  the Feynman diagrams can be divided into  factorizable and nonfactorizable diagrams in the color space,
 the contributions  at the order $\mathcal{O}(\alpha_s^k)$ with $k\leq1$, which are factorizable in the color space, are exactly  canceled out    by the meson-meson scattering states at the hadron side,
the nonfactorizable diagrams, if having  a Landau singularity, begin to make contributions  to the tetraquark molecular states,  the tetraquark molecular states begin to receive contributions at the order $\mathcal{O}(\alpha_s^2)$  \cite{Chu-Sheng-PRD-1,Chu-Sheng-PRD-2}. The assertion is also applicable for the  diquark-antidiquark type currents, as they can be rearranged into the color-singlet-color-singlet type currents through the Fierz re-ordering.

In Ref.\cite{WangZG-Landau}, we  provide   solid  proofs to show that the  Landau equation is of no use  to study  the Feynman diagrams in the QCD sum rules for the tetraquark molecular states and tetraquark states, such as the quarks and gluons are confined objects and thus cannot be put on mass-shell; the operator product expansion is carried out at the region $p^2\ll -\Lambda_{QCD}^2$ rather than at the region $p^2>\Delta^2$ to have Landau singularities; the lowest order Feynman diagrams  have Landau singularities at the region $p^2>\Delta^2$, if not assuming  the factorizable diagrams in the color space only make contributions to the two-meson scattering states, where the $\Delta^2$ denotes  the thresholds. We refute the assertion of Lucha, Melikhov and Sazdjian in details, and  use  two color-singlet-color-singlet type currents as examples to show that the  meson-meson scattering states alone cannot saturate the QCD sum rules,  while
the tetraquark molecular states alone can saturate the QCD sum rules,  the   tetraquark molecular states begin to receive contributions at the order $\mathcal{O}(\alpha_s^0/\alpha_s^1)$ rather than at the order $\mathcal{O}(\alpha_s^2)$ \cite{WangZG-Landau}.
In the light flavor sector,   Lee and Kochelev study  the two-pion  contributions in the QCD sum rules for the scalar meson $f_0(600)$   as the tetraquark state, and observe that the contributions at  the order $\mathcal{O}(\alpha_s^k)$  with $k\leq1$ in the operator product expansion cannot be canceled out by the two-pion  scattering states \cite{Lee-0702-PRD}.

In this article, we extend our previous work to study the $D^*\bar{D}^*$, $D^*D^*$, $D^*D^*\bar{D}^*$ and $D^*D^*D^*$ tetraquark and hexaquark molecular states with the QCD sum rules, and examine whether or not it is necessary to distinguish the factorizable and nonfactorizable properties of the Feynman diagrams in the color space so as to interpret  them in terms of the hadronic observables. Phenomenologically,
the $D^*\bar{D}K$, $D^*\bar{D}^*K$, $D^*D^*\bar{D}$ and $D^*D^*\bar{D}^*$ hexaquark molecular  states  have been studied with the heavy quark spin symmetry \cite{DvDvK-Ren,Valderrama-DvDvDv}, the $BD\bar{D}$, $BDD$, $DB^*\bar{B}^*$, $D^*B^*\bar{B}^*$, $DB\bar{B}$ and $D^*B\bar{B}$ hexaquark molecular states have been studied with the fixed center approximation to the Faddeev equations \cite{BDD-Oset,DvBvBv-Dias}. In Ref.\cite{DiZY-DDvK}, we study the $D\bar{D}^*K$ hexaquark molecular state
with the QCD sum rules by considering the contributions of the vacuum condensates up to dimension-16. In the early works \cite{Nielsen-DvDv-D-6,MQHuang-DvDv-6,JRZhang-DvDv-D-6}, the $D^*\bar{D}^*$ molecular states were studied via the QCD sum rules by carrying out the operator product expansion up to the vacuum condensates of dimension-6, the Borel platforms were  not very flat. In Ref.\cite{Narison-DvDv-D-8}, the $D^*\bar{D}^*$ molecular states are studied via the QCD sum rules by carrying out the operator product expansion up to the vacuum condensates of dimension-8 and partly taking into account the perturbative corrections to the perturbative terms, however, too large continuum threshold parameters and too large energy scales of the QCD spectral densities are taken.  In this article, we study the molecular states of the $D^*$-mesons in a consistent way.

The article is arranged as follows:  we derive the QCD sum rules for the masses and pole residues  of  the
$D^*\bar{D}^*$, $D^*D^*$, $D^*D^*\bar{D}^*$ and $D^*D^*D^*$ tetraquark and hexaquark molecular states, and examine the properties of the Feynman diagrams in Sect.2;  in Sect.3, we present the numerical results and discussions; and Sect.4 is reserved for our
conclusion.

\section{QCD sum rules for the tetraquark and hexaquark molecular states}
Firstly, let us  write down  the two-point correlation functions $\Pi(p)$ and $\Pi_{\mu\nu}(p)$ in the QCD sum rules,
\begin{eqnarray}
\Pi(p)&=&i\int d^4x e^{ip \cdot x} \langle0|T\Big\{J(x)J^{\dagger}(0)\Big\}|0\rangle \, , \nonumber\\
\Pi_{\mu\nu}(p)&=&i\int d^4x e^{ip \cdot x} \langle0|T\Big\{J_{\mu}(x)J^{\dagger}_{\nu}(0)\Big\}|0\rangle \, ,
\end{eqnarray}
where $J(x)=J_{c\bar{c}}(x)$, $ J_{cc}(x)$, $J_{\mu}(x)= J^{cc\bar{c}}_\mu(x)$, $J^{ccc}_\mu(x)$,
 \begin{eqnarray}
 J_{c\bar{c}}(x)&=& \bar{c}(x)\gamma_{\alpha}q(x)\,\bar{q}(x)\gamma^{\alpha}c(x)\, , \nonumber\\
 J_{cc}(x)&=& \bar{q}(x)\gamma_{\alpha}c(x)\,\bar{q}(x)\gamma^{\alpha}c(x)\, , \nonumber\\
 J^{cc\bar{c}}_\mu(x)&=&  \bar{c}(x)\gamma_{\alpha}q(x)\,\bar{q}(x)\gamma^{\alpha}c(x)\,\bar{q}(x)\gamma_{\mu}c(x)\, , \nonumber\\
 J^{ccc}_\mu(x)&=&  \bar{q}(x)\gamma_{\alpha}c(x)\,\bar{q}(x)\gamma^{\alpha}c(x)\,\bar{q}(x)\gamma_{\mu}c(x)\, ,
\end{eqnarray}
where $q=u$ or $d$.
The color-singlet-color-singlet type currents $J_{c\bar{c}}(x)$ and $ J_{cc}(x)$ couple potentially to the scalar hidden-charm and doubly charmed tetraquark molecular states or two-meson scattering states, respectively, the color-singlet-color-singlet-color-singlet type currents $J^{cc\bar{c}}_\mu(x)$ and $J^{ccc}_\mu(x)$ couple potentially to the vector charmed plus hidden-charm and triply charmed hexaquark molecular states or three-meson scattering states, respectively. The currents $J_{c\bar{c}}(x)$, $J_{cc}(x)$, $J^{cc\bar{c}}_\mu(x)$ and $J^{ccc}_\mu(x)$ have the isospins  $I=0$, $1$, $\frac{1}{2}$ and  $\frac{3}{2}$, respectively.
In the isospin limit, the currents with the same isospins couple potentially to the hadrons or hadron-systems with almost degenerated  masses.

At the hadron side of the correlation functions   $\Pi(p)$ and $\Pi_{\mu\nu}(p)$,   we  isolate  the contributions of   the lowest hidden-charm and doubly-charmed tetraquark molecular states  and the charmed plus hidden-charm and triply-charmed hexaquark molecular states to obtain the hadron representation,
\begin{eqnarray}
\Pi(p)&=& \frac{\lambda_{T}^2}{M_{T}^2-p^2}+\cdots=\Pi_{T}(p^2)\, ,   \nonumber\\
\Pi_{\mu\nu}(p)&=& \frac{\lambda_{H}^2}{M_{H}^2-p^2}\left( -g_{\mu\nu}+\frac{p_{\mu}p_{\nu}}{p^2}\right)+\cdots\, ,   \nonumber\\
&=&\Pi_{H}(p^2)\left( -g_{\mu\nu}+\frac{p_{\mu}p_{\nu}}{p^2}\right)+\cdots\, ,
\end{eqnarray}
where the subscripts  $T$ and $H$ denote the scalar $D^*\bar{D}^*$, $D^*D^*$ tetraquark molecular states  and the vector $D^*D^*\bar{D}^*$, $D^*D^*D^*$ hexaquark molecular states, respectively, and we have used the definitions for the pole residues,
\begin{eqnarray}
\langle 0|J(0)|T(p)\rangle &=&\lambda_{T}\, ,\nonumber\\
\langle 0|J_\mu(0)|H(p)\rangle &=&\lambda_{H}\,\varepsilon_\mu\, ,
\end{eqnarray}
  the $\varepsilon_\mu$ are the polarization vectors  of the vector  hexaquark molecular states.
In Ref.\cite{WangZG-Landau}, we observe that, for the color-singlet-color-singlet type currents,  the meson-meson scattering states alone cannot saturate the QCD sum rules, while the tetraquark molecular states alone can saturate the QCD sum rules, the net effects of the two-meson scattering states amount to
modifying the pole residues considerably  without influencing  the predicted masses.  In this article, we only take into account the tetraquark and hexaquark molecular states, as the widths of those molecular states,   which absorb the contributions of the two-meson or three-meson  scattering states, are unknown.

At the QCD side of the correlation functions $\Pi(p)$ and $\Pi_{\mu\nu}(p)$, we contract the $q$ and $c$ quark fields  with the Wick's  theorem, and obtain the results,
\begin{eqnarray}\label{c-barc-CF}
\Pi_{c\bar{c}}(p)&=&i   \int d^4x\, e^{ip\cdot x}\, {\rm Tr}\left[\gamma_\alpha S^{ij}(x)   \gamma_\beta   C^{ji}(-x) \right] {\rm Tr}\left[\gamma^\alpha C^{n m}(x)   \gamma^\beta  S^{mn}(-x)\right]\, ,
\end{eqnarray}

\begin{eqnarray}\label{c-c-CF}
\Pi_{cc}(p)&=&2i   \int d^4x\, e^{ip\cdot x}\,\Big\{ {\rm Tr}\left[\gamma_\alpha C^{ij}(x)   \gamma_\beta   S^{ji}(-x) \right] {\rm Tr}\left[\gamma^\alpha C^{n m}(x)   \gamma^\beta  S^{mn}(-x)\right]  \nonumber \\
&&-{\rm Tr}\left[\gamma_\alpha C^{ij}(x)   \gamma_\beta   S^{jk}(-x)  \gamma^\alpha C^{km}(x)   \gamma^\beta  S^{mi}(-x)\right] \Big\}\, ,
\end{eqnarray}

\begin{eqnarray}\label{cc-barc-CF}
\Pi^{cc\bar{c}}_{\mu\nu}(p)&=&i   \int d^4x\, e^{ip\cdot x}\nonumber\\
 &&\Big\{ -{\rm Tr}\left[\gamma_\alpha S^{ij}(x)   \gamma_\beta   C^{ji}(-x) \right] {\rm Tr}\left[\gamma^\alpha C^{km}(x)   \gamma^\beta   S^{mk}(-x) \right] {\rm Tr}\left[\gamma_\mu C^{ln}(x)   \gamma_\nu  S^{nl}(-x)\right] \nonumber \\
  && -{\rm Tr}\left[\gamma_\alpha S^{ij}(x)   \gamma_\beta   C^{ji}(-x) \right] {\rm Tr}\left[\gamma^\alpha C^{km}(x)   \gamma_\nu   S^{mk}(-x) \right] {\rm Tr}\left[\gamma_\mu C^{ln}(x)   \gamma^\beta  S^{nl}(-x)\right] \nonumber \\
  &&+{\rm Tr}\left[\gamma_\alpha S^{ij}(x)   \gamma_\beta   C^{ji}(-x) \right] {\rm Tr}\left[\gamma^\alpha C^{km}(x)   \gamma^\beta   S^{ml}(-x) \gamma_\mu C^{ln}(x)   \gamma_\nu  S^{nk}(-x)\right] \nonumber \\
   && +{\rm Tr}\left[\gamma_\alpha S^{ij}(x)   \gamma_\beta   C^{ji}(-x) \right] {\rm Tr}\left[\gamma^\alpha C^{km}(x)   \gamma_\nu   S^{ml}(-x) \gamma_\mu C^{ln}(x)   \gamma^\beta  S^{nk}(-x)\right] \Big\}\, ,
\end{eqnarray}

\begin{eqnarray}\label{cc-c-CF}
\Pi^{ccc}_{\mu\nu}(p)&=&2i   \int d^4x\, e^{ip\cdot x}\nonumber\\
 &&\Big\{ -{\rm Tr}\left[\gamma_\alpha C^{ij}(x)   \gamma_\beta   S^{ji}(-x) \right] {\rm Tr}\left[\gamma^\alpha C^{kl}(x)   \gamma^\beta   S^{lk}(-x) \right] {\rm Tr}\left[\gamma_\mu C^{mn}(x)   \gamma_\nu  S^{nm}(-x)\right] \nonumber \\
  && -2{\rm Tr}\left[\gamma_\alpha C^{ij}(x)   \gamma_\beta   S^{ji}(-x) \right] {\rm Tr}\left[\gamma^\alpha C^{kl}(x)   \gamma_\nu   S^{lk}(-x) \right] {\rm Tr}\left[\gamma_\mu C^{mn}(x)   \gamma^\beta  S^{nm}(-x)\right] \nonumber \\
  &&+2{\rm Tr}\left[\gamma_\alpha C^{ij}(x)   \gamma_\beta   S^{ji}(-x) \right] {\rm Tr}\left[\gamma^\alpha C^{kl}(x)   \gamma^\beta   S^{lm}(-x) \gamma_\mu C^{mn}(x)   \gamma_\nu  S^{nk}(-x)\right] \nonumber \\
 &&+2{\rm Tr}\left[\gamma_\alpha C^{ij}(x)   \gamma_\beta   S^{ji}(-x) \right] {\rm Tr}\left[\gamma^\alpha C^{kl}(x)   \gamma_\nu   S^{lm}(-x) \gamma_\mu C^{mn}(x)   \gamma^\beta  S^{nk}(-x)\right] \nonumber \\
 &&+2{\rm Tr}\left[\gamma_\alpha C^{ij}(x)   \gamma_\nu   S^{ji}(-x) \right] {\rm Tr}\left[\gamma^\alpha C^{kl}(x)   \gamma_\beta   S^{lm}(-x) \gamma_\mu C^{mn}(x)   \gamma^\beta  S^{nk}(-x)\right] \nonumber \\
 &&+2{\rm Tr}\left[\gamma_\mu C^{ij}(x)   \gamma_\beta   S^{ji}(-x) \right] {\rm Tr}\left[\gamma_\alpha C^{kl}(x)   \gamma^\beta   S^{lm}(-x) \gamma^\alpha C^{mn}(x)   \gamma_\nu  S^{nk}(-x)\right] \nonumber \\
 &&+{\rm Tr}\left[\gamma_\mu C^{ij}(x)   \gamma_\nu   S^{ji}(-x) \right] {\rm Tr}\left[\gamma_\alpha C^{kl}(x)   \gamma_\beta   S^{lm}(-x) \gamma^\alpha C^{mn}(x)   \gamma^\beta  S^{nk}(-x)\right] \nonumber \\
 && -2{\rm Tr}\left[\gamma_\alpha C^{ij}(x)   \gamma_\beta   S^{jk}(-x)  \gamma^\alpha C^{kl}(x)   \gamma^\beta   S^{lm}(-x)  \gamma_\mu C^{mn}(x)   \gamma_\nu  S^{ni}(-x)\right] \nonumber \\
 && -2{\rm Tr}\left[\gamma_\alpha C^{ij}(x)   \gamma_\beta   S^{jk}(-x)  \gamma^\alpha C^{kl}(x)   \gamma_\nu   S^{lm}(-x)  \gamma_\mu C^{mn}(x)   \gamma^\beta  S^{ni}(-x)\right] \nonumber \\
 && -2{\rm Tr}\left[\gamma_\alpha C^{ij}(x)   \gamma_\nu   S^{jk}(-x)  \gamma^\alpha C^{kl}(x)   \gamma_\beta   S^{lm}(-x)  \gamma_\mu C^{mn}(x)   \gamma^\beta  S^{ni}(-x)\right] \Big\} \, ,
\end{eqnarray}
where the $S_{ij}(x)$ and $C_{ij}(x)$ are the  full light and heavy quark propagators, respectively,
\begin{eqnarray}\label{LQuarkProg}
S_{ij}(x)&=& \frac{i\delta_{ij}\!\not\!{x}}{ 2\pi^2x^4}-\frac{\delta_{ij}\langle
\bar{q}q\rangle}{12} -\frac{\delta_{ij}x^2\langle \bar{q}g_s\sigma Gq\rangle}{192} -\frac{ig_sG^{a}_{\alpha\beta}t^a_{ij}(\!\not\!{x}
\sigma^{\alpha\beta}+\sigma^{\alpha\beta} \!\not\!{x})}{32\pi^2x^2}  \nonumber\\
&& -\frac{\delta_{ij}x^4\langle \bar{q}q \rangle\langle g_s^2 GG\rangle}{27648} -\frac{1}{8}\langle\bar{q}_j\sigma^{\mu\nu}q_i \rangle \sigma_{\mu\nu}+\cdots \, ,
\end{eqnarray}
\begin{eqnarray}\label{HQuarkProg}
C_{ij}(x)&=&\frac{i}{(2\pi)^4}\int d^4k e^{-ik \cdot x} \left\{
\frac{\delta_{ij}}{\!\not\!{k}-m_c}
-\frac{g_sG^n_{\alpha\beta}t^n_{ij}}{4}\frac{\sigma^{\alpha\beta}(\!\not\!{k}+m_c)+(\!\not\!{k}+m_c)
\sigma^{\alpha\beta}}{(k^2-m_c^2)^2}\right.\nonumber\\
&&\left. -\frac{g_s^2 (t^at^b)_{ij} G^a_{\alpha\beta}G^b_{\mu\nu}(f^{\alpha\beta\mu\nu}+f^{\alpha\mu\beta\nu}+f^{\alpha\mu\nu\beta}) }{4(k^2-m_c^2)^5}+\cdots\right\} \, ,\nonumber\\
f^{\alpha\beta\mu\nu}&=&(\!\not\!{k}+m_c)\gamma^\alpha(\!\not\!{k}+m_c)\gamma^\beta(\!\not\!{k}+m_c)\gamma^\mu(\!\not\!{k}+m_c)\gamma^\nu(\!\not\!{k}+m_c)\, ,
\end{eqnarray}
and  $t^n=\frac{\lambda^n}{2}$, the $\lambda^n$ is the Gell-Mann matrix
\cite{Reinders85,WangHuangtao-PRD,Pascual-1984}, we add the subscripts $c\bar{c}$, $cc$, $cc\bar{c}$ and $ccc$ to denote the corresponding currents.
In the full light quark propagator, see Eq.\eqref{LQuarkProg}, we add the term $\langle\bar{q}_j\sigma_{\mu\nu}q_i \rangle$,  which comes  from the  Fierz rearrangement  of the quark-antiquark pair $\langle q_i \bar{q}_j\rangle$ to  absorb the gluons  emitted from other  quark lines  to extract  the mixed condensates   $\langle\bar{q}g_s\sigma G q\rangle$, $\langle\bar{q}g_s\sigma G q\rangle^2$ and $\langle\bar{q}g_s\sigma G q\rangle^3$, respectively \cite{WangHuangtao-PRD}.

In Eqs.\eqref{c-barc-CF}-\eqref{cc-c-CF}, we assume dominance of the vacuum intermediate state tacitly, and insert  the vacuum intermediate
state in all the channels and neglect  the contributions  of all the other states. In the original works, Shifman,  Vainshtein and  Zakharov took
 the factorization hypothesis according to  two reasons \cite{SVZ79}.  One is the rather large value of the quark condensate $\langle\bar{q}q\rangle$, the other is the duality between the quark and physical states, which implies that counting both the quark and
physical states may well become a double counting since they reproduce each other \cite{SVZ79}.

In the QCD sum rules for the traditional mesons, we usually introduce a parameter $\kappa$ to parameterize the  deviation from the factorization hypothesis by hand \cite{Review-rho-kappa}, for example, in the case of the four quark condensate,  $\langle\bar{q}q\rangle^2 \to \kappa\langle\bar{q}q\rangle^2$ \cite{Review-rho-kappa,Narison-rho}. In fact, the  vacuum saturation works well in the large $N_c$ limit \cite{Novikov--shifman}.  As the $\langle\bar{q}q\rangle^2$ is always companied with the fine-structure constant $\alpha_s=\frac{g_s^2}{4\pi}$, and plays a minor important role,
the deviation from $\kappa=1$, for example, $\kappa=2\sim 3$, cannot make much difference, though the value $\kappa>1$ can lead to better QCD sum rules in some cases.

 However, in the QCD sum rules for the tetraquark, pentaquark and hexaquark (molecular) states, the four-quark condensate plays an important role,
a large value, for example, $\kappa=2$, can destroy the platforms in the QCD sum rules for the current $J_{c\bar{c}}(x)$. In calculations, we observe that the optimal value is $\kappa=1$, the vacuum saturation works well in the QCD sum rules for the multiquark states.

Up to now, all the multiquark states are studied with the  procedure illustrated in Eqs.\eqref{c-barc-CF}-\eqref{HQuarkProg}  by assuming the vacuum saturation for the higher dimensional vacuum condensates tacitly in performing the operator product expansion, except for in some case the parameter $\kappa$ is introduced
for the sake of  fine-tuning \cite{Narison-DvDv-D-8}. The true values of the higher dimensional vacuum condensates, even the four quark condensates $\langle \bar{q}\Gamma q \bar{q}\Gamma^\prime q\rangle$, remain unknown or poorly known,  where the $\Gamma$ and $\Gamma^\prime$ stand for the Dirac $\gamma$-matrixes, we cannot obtain robust estimations  about the effects beyond the vacuum saturation.

In the QCD sum rules for the multiquark states, such as the tetraquark, pentaquark and hexaquark (molecular) states,  we often encounter the vacuum condensates $\langle \bar{q}q\rangle^2$ and $\langle \bar{q}q\rangle^3$, we  have the choice to introduce the parameter $\kappa$ to parameterize the  deviation from the factorization hypothesis, although the most commonly used  value $\kappa=1$ works well. If the true value $\kappa >1$ or $\gg 1$,  the QCD sum rules for the tetraquark, pentaquark and hexaquark (molecular) states have considerably larger systematic uncertainties and are
less reliable than those of traditional mesons and baryons \cite{Gubler-SB}. In this respect, we make predictions for the multiquark masses with the QCD sum rules based on the  vacuum saturation (i.e. $\kappa=1$), then confront them to the experimental data in the future to test the theoretical calculations. We can also choose the value $\kappa > 1$ and obtain predictions to be compared with the  experimental data in the future. Many works are still needed to obtain the pertinent value of the $\kappa$.

 We usually neglect the tadpole-like contributions from the contractions of the light quarks in the same currents in the QCD sum rules for the multiquark states. For the currents $J_{c\bar{c}}(x)$ and $J^{cc\bar{c}}_\mu(x)$, if we contract the light quarks in the same currents, we can obtain the following  tadpole-like contributions,
 \begin{eqnarray}
 J_{c\bar{c}}(x)& \to& \hat{J}_{c\bar{c}}(x)=-\frac{\langle\bar{q}q\rangle}{3}\bar{c}(x) c(x)\, , \nonumber\\
 J^{cc\bar{c}}_\mu(x)&\to&\hat{J}^{cc\bar{c}}_\mu(x)= -\frac{5\langle\bar{q}q\rangle}{12} \bar{c}(x) c(x)\,\bar{q}(x)\gamma_{\mu}c(x)-\frac{i\langle\bar{q}q\rangle}{12} \bar{c}(x)\sigma_{\mu\alpha} c(x)\,\bar{q}(x)\gamma^{\alpha}c(x)\, .
\end{eqnarray}
The induced currents $\hat{J}_{c\bar{c}}(x)$ and $\hat{J}^{cc\bar{c}}_\mu(x)$ have two and four valence quarks, respectively, and they are  irrelevant in the present case and can be neglected safely.

Now let us compute  the integrals both in the coordinate space and momentum space in Eqs.\eqref{c-barc-CF}-\eqref{cc-c-CF} to obtain the correlation functions $\Pi_{T}(p^2)$ and $\Pi_{H}(p^2)$   at the
quark-gluon  level, and then  obtain the QCD spectral densities through   dispersion relation,
\begin{eqnarray}\label{QCD-rho1-2}
 \rho_{T/H,QCD}(s) &=&{\rm lim}_{\epsilon\to0}\frac{{\rm Im}\Pi_{T/H}(s+i\epsilon)}{\pi}\, ,
\end{eqnarray}
the QCD spectral densities $ \rho_{T,QCD}(s)$ are given explicitly in the appendix, while the analytical expressions of the QCD spectral densities $ \rho_{H,QCD}(s)$  are too cumbersome, the interested readers can contract me via E-mail to obtain them.

\begin{figure}
 \centering
  \includegraphics[totalheight=3cm,width=3cm]{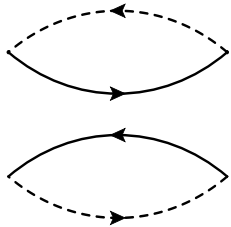}
 \caption{ The  Feynman diagram  for the lowest order  contribution for the current $J_{c\bar{c}}(x)$, where the solid lines and dashed lines represent  the light quarks and heavy quarks, respectively.  }\label{DvDv-tree}
\end{figure}

\begin{figure}
 \centering
  \includegraphics[totalheight=3cm,width=12cm]{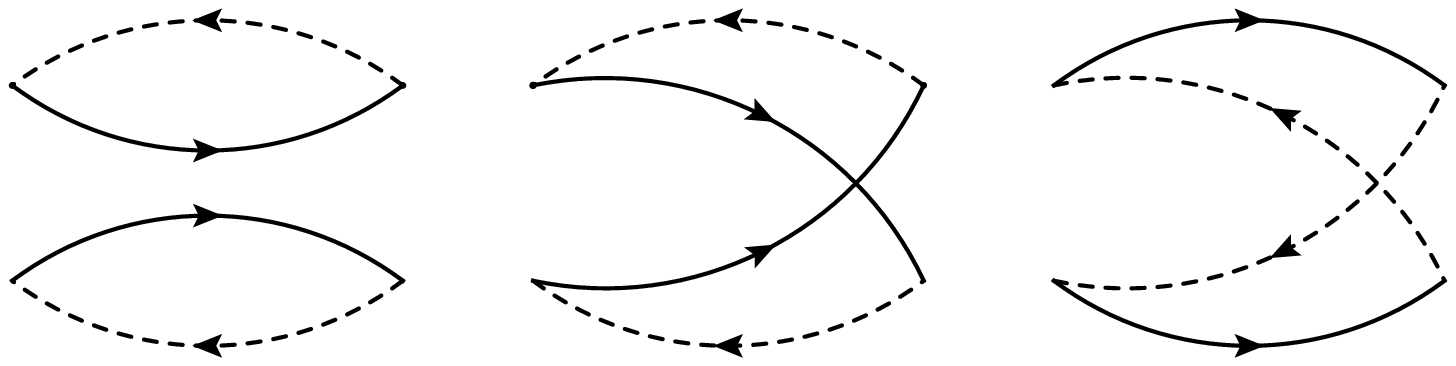}
 \caption{ The  Feynman diagrams  for the lowest order  contributions for the current $J_{cc}(x)$, where the solid lines and dashed lines represent  the light quarks and heavy quarks, respectively.  }\label{DvDv-cc-tree}
\end{figure}

\begin{figure}
 \centering
  \includegraphics[totalheight=5cm,width=12cm]{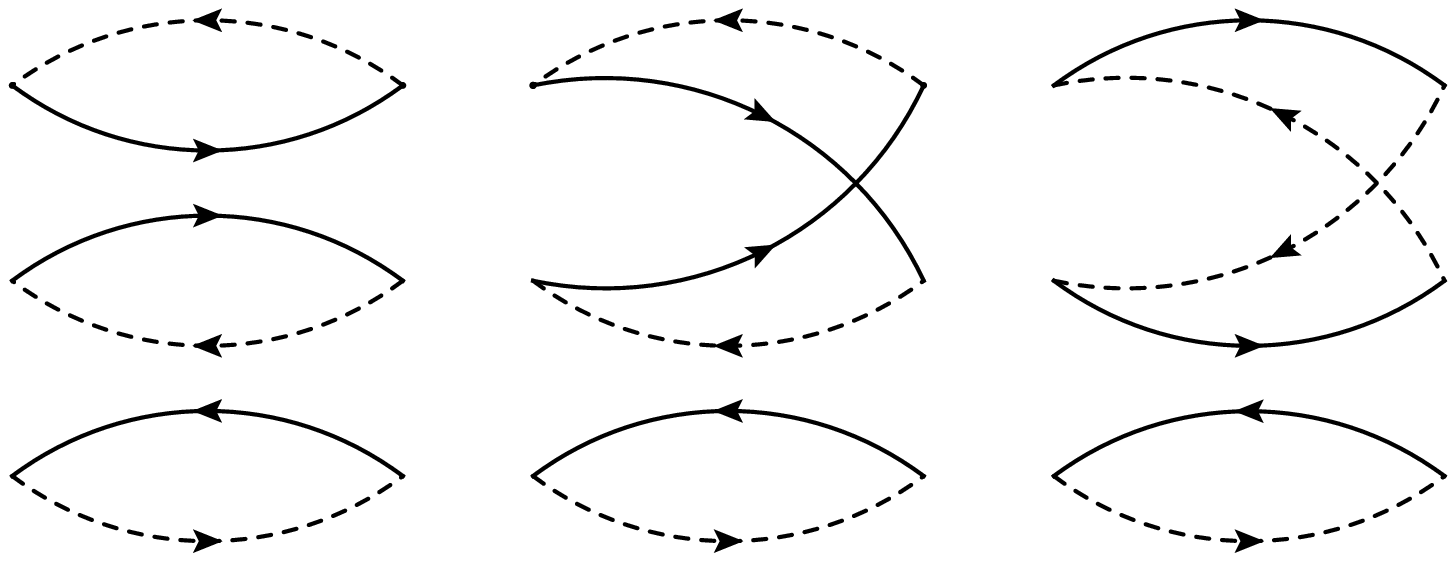}
 \caption{ The  Feynman diagrams  for the lowest order  contributions for the current $J_\mu^{cc\bar{c}}(x)$, where the solid lines and dashed lines represent  the light quarks and heavy quarks, respectively.  }\label{DvDvDv-tree}
\end{figure}

\begin{figure}
 \centering
  \includegraphics[totalheight=5cm,width=12cm]{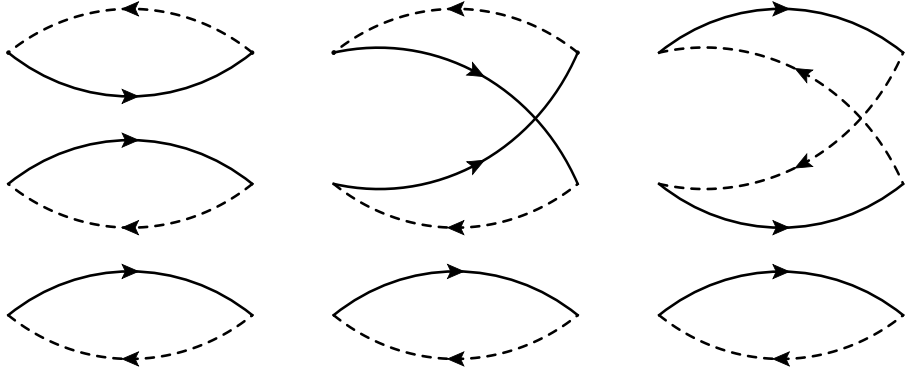}
 \caption{ The factorizable Feynman diagrams in the color space for the lowest order  contributions for the current $J_\mu^{ccc}(x)$, where the solid lines and dashed lines represent  the light quarks and heavy quarks, respectively.  }\label{DvDvDv-ccc-tree-fact}
\end{figure}

\begin{figure}
 \centering
  \includegraphics[totalheight=6cm,width=7cm]{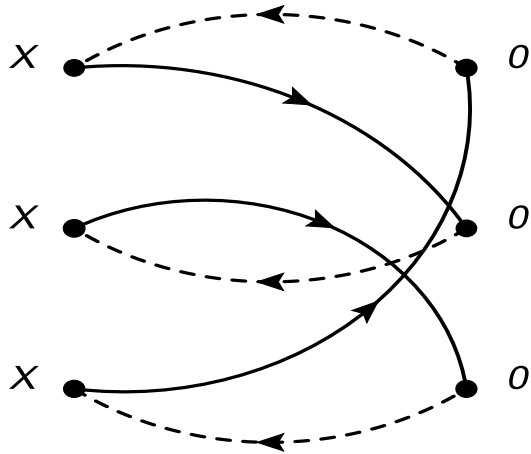}
 \caption{ The nonfactorizable Feynman diagrams in the color space for the lowest order  contributions for the current $J_{ccc}(x)$, where the solid lines and dashed lines represent  the light quarks and heavy quarks, respectively, the other diagrams obtained by interchanging of the three vertexes at the point $0$ or $x$ are implied. }\label{DvDvDv-ccc-tree-nonfact}
\end{figure}

All the contributions in the operator product expansion can be shown explicitly using the Feynman diagrams.
  In the Feynman diagrams drawn in Figs.\ref{DvDv-tree}-\ref{DvDvDv-ccc-tree-nonfact}, we show the lowest order contributions.
   If we substitute the light and heavy quark lines with the full light and heavy quark propagators in Eqs.\eqref{LQuarkProg}-\eqref{HQuarkProg}, respectively,
 we can obtain all the Feynman diagrams.
   From the figures, we can see that there are only factorizable contributions in the color space for the currents $J_{c\bar{c}}(x)$ and $J_\mu^{cc\bar{c}}(x)$, while there are both  factorizable and  nonfactorizable contributions in the color space for the currents $J_{cc}(x)$ and $J_\mu^{ccc}(x)$. There are also nonfactorizable sub-clusters in the Feynman diagrams for the current $J_\mu^{cc\bar{c}}(x)$, see the second and the third diagrams in Fig.\ref{DvDvDv-tree}.
Thereafter, we will use the nomenclatures  "factorizable contributions" and  "nonfactorizable contributions" to denote the contributions come from the
factorizable and nonfactorizable Feynman diagrams in the color space, respectively.

From the factorizable  diagrams, see Fig.\ref{DvDv-tree} and Figs.\ref{DvDvDv-tree}-\ref{DvDvDv-ccc-tree-fact}, we can obtain both the factorizable and nonfactorizable  diagrams, while from the nonfactorizable  diagrams, see Fig.\ref{DvDv-cc-tree} and Fig.\ref{DvDvDv-ccc-tree-nonfact}, we can obtain only the nonfactorizable  diagrams.
Lucha, Melikhov and Sazdjian assert that those nonfactorizable Feynman diagrams shown in Fig.\ref{DvDv-cc-tree} and Fig.\ref{DvDvDv-ccc-tree-nonfact}  can be deformed into the box diagrams \cite{Chu-Sheng-PRD-2}. It is unfeasible, those nonfactorizable Feynman diagrams can only be deformed into the box diagrams in the color space, not in the momentum space.

\begin{figure}
 \centering
  \includegraphics[totalheight=3cm,width=15cm]{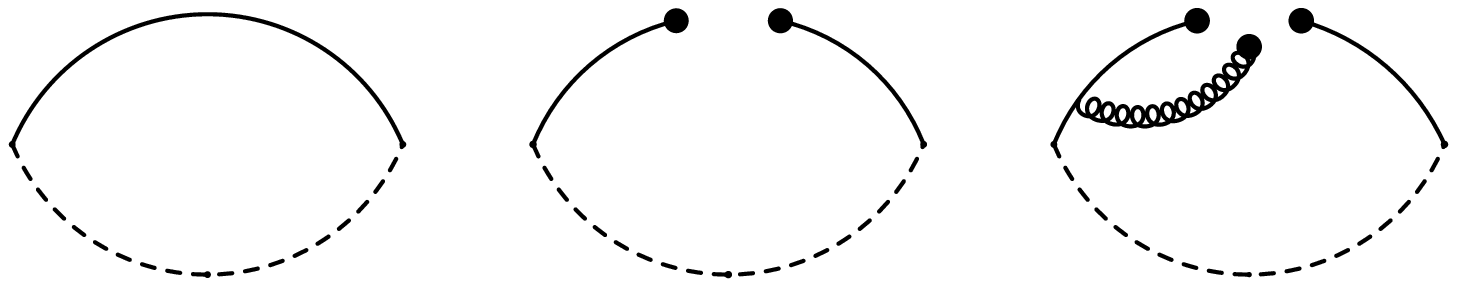}
 \caption{ The typical Feynman diagrams which can  be factorized into  two colored quark lines  for the conventional heavy mesons, where the solid line and dashed line denote the  light  quark and heavy quark, respectively.}\label{Two-quark-fact}
\end{figure}

\begin{figure}
 \centering
  \includegraphics[totalheight=3cm,width=15cm]{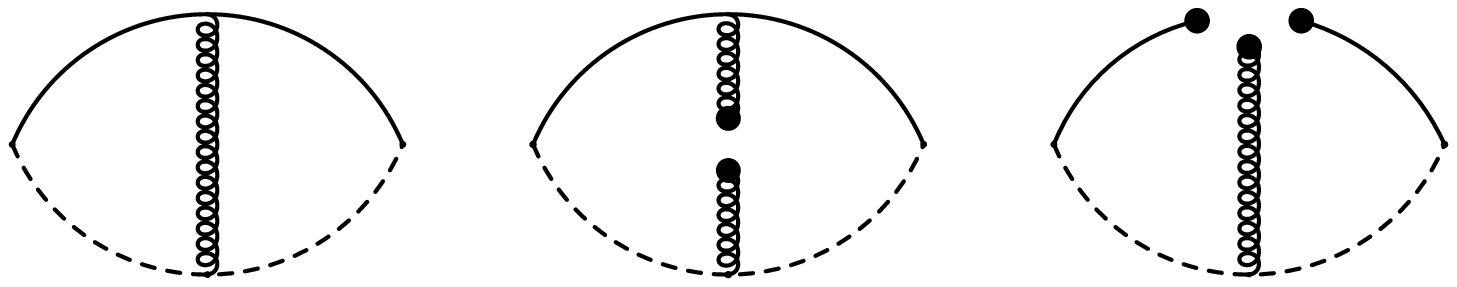}
 \caption{ The typical Feynman diagrams which cannot  be factorized into  two colored quark lines  for the conventional heavy mesons, where the solid line and dashed line denote the  light  quark and heavy quark, respectively.}\label{Two-quark-nonfact}
\end{figure}

According to the assertion of  Lucha, Melikhov and Sazdjian \cite{Chu-Sheng-PRD-1,Chu-Sheng-PRD-2},  the factorizable (disconnected) diagrams in the color space only make contributions  to the  meson-meson scattering states. From the lowest Feynman diagrams shown in Figs.\ref{DvDv-tree}-\ref{DvDvDv-ccc-tree-nonfact}, we can draw the conclusion tentatively that we can obtain more good QCD sum rules from the currents $J_{cc}(x)$ and $J^{ccc}_\mu(x)$ than from the currents $J_{c\bar{c}}(x)$ and $J^{cc\bar{c}}_\mu(x)$, as there are connected (nonfactorizable) diagrams besides disconnected (factorizable) diagrams.

In fact, it is useless  to  distinguish the factorizable and nonfactorizable properties of the Feynman diagrams in the operator product expansion, where the
short-distance contributions above a certain energy scale $\mu$ are included in the Wilson's coefficients, the long-distance contributions below the special energy scale $\mu$ are included in the vacuum condensates. In general, we can choose any energy scales at which the perturbative QCD calculations are feasible.
 Besides the uncertainties originate from the energy scales, additional uncertainties come from the fact that
it is impossible to separate  the soft tails in the quark-loop diagrams using the standard  Feynman-diagram technique.
We cannot obtain the information asserted by Lucha, Melikhov and Sazdjian  from the Feynman diagrams  in the operator product expansion \cite{Chu-Sheng-PRD-1,Chu-Sheng-PRD-2},
 we can only  obtain information about the short-distance and long-distance contributions.

 We can borrow some ideas from the QCD sum rules for the conventional heavy mesons,  in which the hadronic spectral densities can be written as,
   \begin{eqnarray}
\rho_H(s)&=&f_H^2m_H^2\delta(s-m_H^2)+\rho_{QCD}(s)\Theta(s-s_0)\, ,
 \end{eqnarray}
where the subscript $H$ denotes the $D$ and $B$ mesons, the $f_{H}$ are the decay constants, the hadronic spectral densities above the continuum thresholds $s_0$ are approximated by the perturbative contributions as only the perturbative contributions are left. In the operator product expansion, we often encounter  the Feynman diagrams  shown in Figs.\ref{Two-quark-fact}-\ref{Two-quark-nonfact},
the Feynman diagrams shown in Fig.\ref{Two-quark-fact} (Fig.\ref{Two-quark-nonfact}) can be (cannot be) factorized into two colored quark lines. Analogously, could we assert that the  Feynman diagrams shown in Fig.\ref{Two-quark-fact} can be exactly  canceled out by two asymptotic quarks, only
 the Feynman diagrams shown in Fig.\ref{Two-quark-nonfact} make contributions to the heavy mesons? In Ref.\cite{Chu-Sheng-Heavy-meson}, Lucha,  Melikhov and  Simula take into account all those  Feynman diagrams, which is in contrast to the assertion of Lucha, Melikhov and Sazdjian in Refs.\cite{Chu-Sheng-PRD-1,Chu-Sheng-PRD-2}.

In the QCD sum rules for the tetraquark (molecular) states, pentaquark (molecular) states and hexaquark states (or dibaryon states), we take into account the vacuum condensates,  which are vacuum expectations of the quark-gluon operators of the order $\mathcal{O}(\alpha_s^k)$ with $k\leq1$ in a consistent way \cite{WangZG-4-quark-mole,WangZG-Landau,WangHuangtao-PRD,Wang-tetra-formula,ZGWang-CTP,WangZG-IJMPA-2020,WZG-ccc-dibaryon,WZG-hexaquark-ccc}.

There are two light quark lines and two heavy quark lines  in the Feynman diagrams in Figs.\ref{DvDv-tree}-\ref{DvDv-cc-tree},
while there are three light quark lines  and three heavy quark lines in Figs.\ref{DvDvDv-tree}-\ref{DvDvDv-ccc-tree-nonfact},
 if each heavy quark line emits a gluon and each light quark line contributes quark-antiquark pair, we obtain the  quark-gluon  operators $g_sG_{\mu\nu}g_sG_{\alpha\beta}\bar{q}q\bar{q}q$ and $g_sG_{\mu\nu}g_sG_{\alpha\beta}g_sG_{\lambda\tau}\bar{q}q\bar{q}q\bar{q}q$ from the Figs.\ref{DvDv-tree}-\ref{DvDv-cc-tree} and Figs.\ref{DvDvDv-tree}-\ref{DvDvDv-ccc-tree-nonfact}, respectively,  which are of dimension 10 and 15, respectively.
The operator $g_sG_{\mu\nu}g_sG_{\alpha\beta}\bar{q}q\bar{q}q$ leads to the vacuum condensates
$\langle\frac{\alpha_sGG}{\pi}\rangle\langle\bar{q}q\rangle^2$ and $\langle\bar{q}g_s\sigma Gq\rangle^2$, while the operator $g_sG_{\mu\nu}g_sG_{\alpha\beta}g_sG_{\lambda\tau}\bar{q}q\bar{q}q\bar{q}q$ leads to the vacuum condensates
$\langle\frac{\alpha_sGG}{\pi}\rangle\langle\bar{q}q\rangle^2\langle\bar{q}g_s\sigma Gq\rangle$, $\langle g_s^3 GGG\rangle\langle\bar{q}q\rangle^3$ and $\langle\bar{q}g_s\sigma Gq\rangle^3$.

In the present case, if taking the truncation $k\leq1$ for the quark-gluon operators, for the correlation function $\Pi(p)$, the highest dimensional vacuum condensates are $\langle\bar{q}q\rangle^2\langle\frac{\alpha_sGG}{\pi}\rangle$ and $\langle\bar{q}g_s\sigma Gq\rangle^2$, which are of dimension 10, while for the correlation function $\Pi_{\mu\nu}(p)$, the highest dimensional vacuum condensates are $\langle\bar{q}q\rangle^3\langle\frac{\alpha_sGG}{\pi}\rangle$ and $\langle\bar{q} q\rangle\langle\bar{q}g_s\sigma Gq\rangle^2$, which are of dimension 13.   The vacuum condensates $\langle\frac{\alpha_sGG}{\pi}\rangle\langle\bar{q}q\rangle^2\langle\bar{q}g_s\sigma Gq\rangle$, $\langle g_s^3 GGG\rangle\langle\bar{q}q\rangle^3$ and $\langle\bar{q}g_s\sigma Gq\rangle^3$, which are of dimension 15,  come from the quark-gluon operators of the order $\mathcal{O}(\alpha_s^{\frac{3}{2}})$ and should be discarded. In the correlation function $\Pi_{\mu\nu}(p)$, we take into account the vacuum condensate $\langle\bar{q}g_s\sigma Gq\rangle^3$, although it is beyond the truncation $k\leq1$,    and neglect the vacuum condensates $\langle g_s^3 GGG\rangle\langle\bar{q}q\rangle^3$ and $\langle\frac{\alpha_sGG}{\pi}\rangle\langle\bar{q}q\rangle^2\langle\bar{q}g_s\sigma Gq\rangle$ due to their  small values, just like in the QCD sum rules for the
triply charmed dibaryon states and diquark-diquark-diquark type hexaquark states \cite{WZG-ccc-dibaryon,WZG-hexaquark-ccc,WZG-QQQ-baryon}.

In summary,  we carry out the
operator product expansion up to the vacuum condensates of  dimension-10 and dimension-15 for the correlations functions $\Pi(p)$ and $\Pi_{\mu\nu}(p)$ respectively in
a  consistent way.
 For the correlation  function $\Pi(p)$, we  take into account the vacuum condensates $\langle\bar{q}q\rangle$, $\langle\frac{\alpha_sGG}{\pi}\rangle$, $\langle\bar{q}g_s\sigma Gq\rangle$, $\langle\bar{q}q\rangle^2$,  $\langle\bar{q}q\rangle\langle\frac{\alpha_sGG}{\pi}\rangle$,
  $\langle\bar{q} q\rangle\langle\bar{q}g_s\sigma Gq\rangle$,  $\langle\bar{q}q\rangle^2\langle\frac{\alpha_sGG}{\pi}\rangle$ and
   $\langle\bar{q}g_s\sigma Gq\rangle^2$ \cite{WangZG-4-quark-mole,WangZG-Landau,WangHuangtao-PRD,Wang-tetra-formula,ZGWang-CTP}.
  For the correlation  function $\Pi_{\mu\nu}(p)$,  we take into account the vacuum condensates $\langle\bar{q}q\rangle$, $\langle\frac{\alpha_sGG}{\pi}\rangle$, $\langle\bar{q}g_s\sigma Gq\rangle$, $\langle\bar{q}q\rangle^2$,  $\langle\bar{q}q\rangle\langle\frac{\alpha_sGG}{\pi}\rangle$,
  $\langle\bar{q} q\rangle\langle\bar{q}g_s\sigma Gq\rangle$, $\langle\bar{q} q\rangle^3$, $\langle\bar{q}q\rangle^2\langle\frac{\alpha_sGG}{\pi}\rangle$,
   $\langle\bar{q}g_s\sigma Gq\rangle^2$,  $\langle\bar{q} q\rangle^2\langle\bar{q}g_s\sigma Gq\rangle$, $\langle\bar{q}q\rangle^3\langle\frac{\alpha_sGG}{\pi}\rangle$, $\langle\bar{q} q\rangle\langle\bar{q}g_s\sigma Gq\rangle^2$,
   $\langle\bar{q}g_s\sigma Gq\rangle^3$ \cite{WZG-ccc-dibaryon,WZG-hexaquark-ccc}.

Once  the analytical expressions of the QCD spectral densities are obtained,   we match the hadron side with the QCD side of the correlation functions $\Pi_{T}(p^2)$ and $\Pi_{H}(p^2)$  below the continuum threshold  $s_0$ and perform the Borel transform   in regard to $P^2=-p^2$ to obtain  the   QCD sum rules:
\begin{eqnarray}\label{QCDSR-A}
\lambda^{2}_{T/H}\exp\left( -\frac{M_{T/H}^2}{T^2}\right)&=& \int_{\Delta^2}^{s_0}ds \,\rho_{T/H,QCD}(s)\,\exp\left( -\frac{s}{T^2}\right)\, ,
\end{eqnarray}
where the thresholds $\Delta^2=4m_c^2$ and $9m_c^2$ for the
QCD spectral densities $\rho_{T,QCD}(s)$ and $\rho_{H,QCD}(s)$, respectively.

We differentiate  Eq.\eqref{QCDSR-A} in regard to  $\tau=\frac{1}{T^2}$, then eliminate the
 pole residues $\lambda_{T/H}$  and obtain the QCD sum rules for
 the masses   of the scalar $D^*\bar{D}^*$, $D^*D^*$ tetraquark molecular states and   the vector  $D^*D^*\bar{D}^*$, $D^*D^*D^*$ hexaquark molecular states,
 \begin{eqnarray}\label{QCDSR-A-Deri}
 M^2_{T/H} &=& \frac{-\frac{d}{d\tau}\int_{\Delta^2}^{s_0}ds \,\rho_{T/H,QCD}(s)\,\exp\left( -s\tau\right)}{\int_{\Delta^2}^{s_0}ds \,\rho_{T/H,QCD}(s)\,\exp\left( -s\tau\right)}\, .
 \end{eqnarray}

\section{Numerical results and discussions}
At the QCD side, we choose  the standard values of the vacuum condensates $\langle
\bar{q}q \rangle=-(0.24\pm 0.01\, \rm{GeV})^3$,   $\langle
\bar{q}g_s\sigma G q \rangle=m_0^2\langle \bar{q}q \rangle$,
$m_0^2=(0.8 \pm 0.1)\,\rm{GeV}^2$,  $\langle \frac{\alpha_s
GG}{\pi}\rangle=(0.33\,\rm{GeV})^4 $    at the energy scale  $\mu=1\, \rm{GeV}$
\cite{SVZ79,Reinders85,Colangelo-Review}, and choose the $\overline{MS}$ mass  $m_{c}(m_c)=(1.275\pm0.025)\,\rm{GeV}$
 from the Particle Data Group \cite{PDG}, and set the  small $u$ and $d$ quark masses to be zero.
 Furthermore, we take into account the energy-scale dependence of  those  input parameters,
\begin{eqnarray}
\langle\bar{q}q \rangle(\mu)&=&\langle\bar{q}q \rangle({\rm 1GeV})\left[\frac{\alpha_{s}({\rm 1GeV})}{\alpha_{s}(\mu)}\right]^{\frac{12}{25}}\, ,\nonumber\\
\langle\bar{q}g_s \sigma Gq \rangle(\mu)&=&\langle\bar{q}g_s \sigma Gq \rangle({\rm 1GeV})\left[\frac{\alpha_{s}({\rm 1GeV})}{\alpha_{s}(\mu)}\right]^{\frac{2}{25}}\, , \nonumber\\
m_c(\mu)&=&m_c(m_c)\left[\frac{\alpha_{s}(\mu)}{\alpha_{s}(m_c)}\right]^{\frac{12}{25}} \, ,\nonumber\\
\alpha_s(\mu)&=&\frac{1}{b_0t}\left[1-\frac{b_1}{b_0^2}\frac{\log t}{t} +\frac{b_1^2(\log^2{t}-\log{t}-1)+b_0b_2}{b_0^4t^2}\right]\, ,
\end{eqnarray}
  where $t=\log \frac{\mu^2}{\Lambda^2}$, $b_0=\frac{33-2n_f}{12\pi}$, $b_1=\frac{153-19n_f}{24\pi^2}$, $b_2=\frac{2857-\frac{5033}{9}n_f+\frac{325}{27}n_f^2}{128\pi^3}$,  $\Lambda=210\,\rm{MeV}$, $292\,\rm{MeV}$  and  $332\,\rm{MeV}$ for the flavors  $n_f=5$, $4$ and $3$, respectively  \cite{PDG,Narison-mix}, and evolve all the input parameters to the pertinent  energy scales $\mu$  to extract the  masses   of the scalar $D^*\bar{D}^*$, $D^*D^*$ tetraquark molecular states and   the vector  $D^*D^*\bar{D}^*$, $D^*D^*D^*$ hexaquark molecular states with the flavor $n_f=4$, as we cannot obtain energy scale independent QCD sum rules.

The correlation functions $\Pi_{T/H}(p^2)$ can be written as
\begin{eqnarray}
\Pi_{T/H}(p^2)&=&\int_{4/9m^2_c(\mu)}^{s_0} ds \frac{\rho_{QCD}(s,\mu)}{s-p^2}+\int_{s_0}^\infty ds \frac{\rho_{QCD}(s,\mu)}{s-p^2} \, ,
\end{eqnarray}
through dispersion relation at the QCD side, and they are scale independent or independent on the energy scale we choose to carry out the operator product expansion,
\begin{eqnarray}
\frac{d}{d\mu}\Pi_{T/H}(p^2)&=&0\, ,
\end{eqnarray}
which does not mean
\begin{eqnarray}
\frac{d}{d\mu}\int_{4/9m^2_c(\mu)}^{s_0} ds \frac{\rho_{QCD}(s,\mu)}{s-p^2}\rightarrow 0 \, ,
\end{eqnarray}
 due to the two features inherited from the QCD sum rules:\\
$\bullet$ Perturbative corrections are neglected, even in the QCD sum rules for the traditional mesons, we cannot take into account the perturbative corrections up to arbitrary  orders;  the higher dimensional vacuum condensates are factorized into lower dimensional ones based on the vacuum saturation, therefore  the energy scale dependence of the higher dimensional vacuum condensates is modified;\\
$\bullet$ Truncations $s_0$ set in, the correlation between the threshold $4/9m^2_c(\mu)$ and continuum threshold $s_0$ is unknown. \\
After performing the Borel transform, we obtain the integrals
 \begin{eqnarray}
 \int_{4/9m_c^2(\mu)}^{s_0} ds \rho_{QCD}(s,\mu)\exp\left(-\frac{s}{T^2} \right)\, ,
 \end{eqnarray}
which are sensitive to the $c$-quark mass $m_c(\mu)$ or the energy scale $\mu$.  Variations of the energy scale $\mu$ can lead to changes of integral ranges $4/9m_c^2(\mu)-s_0$ of the variable
$ds$ besides the QCD spectral densities $\rho_{QCD}(s,\mu)$, therefore changes of the Borel windows and predicted masses and pole residues.

Although we cannot obtain the QCD sum rules independent on the energy scales of the QCD spectral densities,  we have an energy scale formula to determine the pertinent energy scales consistently.
In this  article, we study the color-singlet-color-singlet type  tetraquark molecular states and color-singlet-color-singlet-color-singlet type  hexaquark molecular states, which have  two charm quarks and three  charm quarks, respectively.
Such two-$c$-quark and three-$c$-quark systems  are characterized by the effective charmed quark mass  or constituent quark mass ${\mathbb{M}}_c$
and the virtuality  $V\sim\sqrt{M^2_{T}-(2{\mathbb{M}}_c)^2}$ or $\sqrt{M^2_{H}-(3{\mathbb{M}}_c)^2}$. We set the energy  scales of the QCD spectral densities to be $\mu\sim V$, and obtain the energy scale formula,
\begin{eqnarray}\label{formula}
\mu&=&\sqrt{M^2_{T}-(2{\mathbb{M}}_c)^2}\, , \nonumber\\
     &=&\sqrt{M^2_{H}-(3{\mathbb{M}}_c)^2}\, ,
\end{eqnarray}
 for the  tetraquark molecular states and hexaquark molecular states, respectively \cite{WangZG-4-quark-mole,WangZG-Landau,WangHuangtao-PRD,Wang-tetra-formula,ZGWang-CTP,WangZG-IJMPA-2020,WZG-ccc-dibaryon,WZG-hexaquark-ccc}.
 Analysis of the $J/\psi$ and $\Upsilon$
with the famous   Cornell potential or Coulomb-potential-plus-linear-potential leads to the constituent quark masses $m_c=1.84\,\rm{GeV}$ and $m_b=5.17\,\rm{GeV}$ \cite{Cornell}, we can set the effective $c$-quark mass equal to the constituent quark mass ${\mathbb{M}}_c=m_c=1.84\,\rm{GeV}$.   The old value ${\mathbb{M}}_c=1.84\,\rm{GeV}$ and updated value  ${\mathbb{M}}_c=1.85\,\rm{GeV}$ fitted in the QCD sum rules for the hidden-charm tetraquark molecular states
 are all consistent with the  constituent quark mass $m_c=1.84\,\rm{GeV}$ \cite{WangZG-4-quark-mole,WangZG-CPC-Y4390}. We can choose the value ${\mathbb{M}}_c=1.84\pm0.01\,\rm{GeV}$ \cite{WangZG-Landau}, take the energy scale formula $\mu=\sqrt{M^2_{T}-(2{\mathbb{M}}_c)^2}$ and
    $\sqrt{M^2_{H}-(3{\mathbb{M}}_c)^2}$ to improve the convergence of the operator product expansion and enhance the pole contributions. It is a remarkable advantage of the present work.

We can rewrite the energy scale formula in the following form,
\begin{eqnarray}\label{formula-Regge}
M^2_{T/H}&=&\mu^2+{\rm Constants}\, ,
\end{eqnarray}
where the Constants have the values $4{\mathbb{M}}_c^2$ or $9{\mathbb{M}}_c^2$.  As we cannot obtain energy scale independent QCD sum rules, we conjecture that the predicted multiquark masses and the pertinent energy scales of the QCD spectral densities have a
 Regge-trajectory-like relation, see  Eq.\eqref{formula-Regge}, where the Constants are free parameters and fitted by the QCD sum rules. Direct calculations have proven that the Constants  have universal values and work well for all the tetraquark and hexaquark molecular states. At the beginning, we do not know the values of the mulitquark masses, we choose the energy scale $\mu=1.0\,\rm{GeV}$ tentatively, then optimize the continuum threshold parameters and Borel parameters to obtain the  predictions  $M_{T/H}$, if they do not satisfy the relation $M^2_{T/H}=\mu^2+{\rm Constants}$, then we set the energy scale $\mu=1.1\,\rm{GeV}$, $1.2\,\rm{GeV}$, $1.3\,\rm{GeV}$, $\cdots$ tentatively, and repeat
the same routine until reach the satisfactory results.

Such a routine can be referred to as trial and error, we search for the best continuum threshold parameters $s_0$ and Borel parameters $T^2$ via trial and error to satisfy the two basic criteria of the QCD sum rules, the one criterion is pole dominance at the hadron  side, the other criterion is
convergence of the operator product expansion at the QCD side.
Firstly, let us  define the pole contributions $\rm{PC}$,
\begin{eqnarray}
{\rm PC}&=& \frac{ \int_{\Delta^2}^{s_0} ds\,\rho_{QCD}(s)\,\exp\left(-\frac{s}{T^2}\right)}{\int_{\Delta^2}^{\infty} ds \,\rho_{QCD}(s)\,\exp\left(-\frac{s}{T^2}\right)}\, ,
\end{eqnarray}
and  define the contributions of the vacuum condensates of dimension  $n$,
  \begin{eqnarray}
D(n)&=& \frac{  \int_{\Delta^2}^{s_0} ds\,\rho_{QCD;n}(s)\,\exp\left(-\frac{s}{T^2}\right)}{\int_{\Delta^2}^{s_0} ds \,\rho_{QCD}(s)\,\exp\left(-\frac{s}{T^2}\right)}\, ,
\end{eqnarray}
where the $\rho_{QCD;n}(s)$ are the QCD spectral densities  containing the vacuum condensates of dimension $n$.
For the hexaquark (molecular) states, the largest power $\rho_{H,QCD}(s)\propto s^7$,
while for the tetraquark (molecular) states, the largest power $\rho_{T,QCD}(s)\propto s^4$,
it is very difficult to satisfy the two  basic criteria of the QCD sum rules  simultaneously, we have to resort to some methods to improve the convergent behaviors  of the operator product expansion and  enhance the pole contributions, the energy scale formula does the work.

We often consult the experimental data to estimate the continuum threshold parameters $s_0$, for the conventional  quark-antiquark-type  or normal  mesons, we can take any  values  satisfy the relation $M_{gr}<\sqrt{s_0}\leq M_{gr}+\overline{\Delta}$, where the subscript $gr$  represents  the ground states,
as there exists  an energy gap $\overline{\Delta}$ between the ground state and the first radial excited state. For the conventional S-wave quark-antiquark-type  mesons, the energy gaps  $\overline{\Delta}$ vary from $m_{m_{K^*(1410)}}-m_{K^*(892)}=522\,\rm{MeV}$ to $m_{\pi(1300)}-m_{\pi}=1160\,\rm{MeV}$, i.e. $\overline{\Delta}=522\sim 1160\,\rm{MeV}$ from the Particle Data Group \cite{PDG}. If we assign the doublet $(D(2550),D^*(2600))$ to be  the first radial excited state of the doublet $(D,D^*)$ \cite{WZG-D2550-2600}, the energy gap between the $D^*$ and $D^*(2600)$ is about  $0.61\,\rm{GeV}$ \cite{PDG}.  In Ref.\cite{WZG-decay-constants}, we study the masses and decay constants of the heavy mesons with the QCD sum rules in a comprehensive way, in calculations, we observe that the continuum threshold parameter $s_0=6.4\pm0.5\,\rm{GeV}^2$ or $\sqrt{s_0}=2.53\pm0.10\,\rm{GeV}$ can lead to satisfactory result  for the vector  meson $D^*$. We usually choose the continuum threshold parameters as  $\sqrt{s_0}=M_{gr}+(0.4\sim0.7)\,\rm{GeV}$
in the QCD sum rules for the conventional quark-antiquark-type  or normal  mesons  \cite{Colangelo-Review}.
In the present work, we study the $D^*\bar{D}^*$, $D^*D^*$ tetraquark molecular states and the $D^*D^*\bar{D}^*$, $D^*D^*D^*$ tetraquark molecular states, it is reasonable to choose the continuum threshold parameters  as $\sqrt{s_0}=M_{T/H}+0.55\,\rm{GeV}\pm0.10\,\rm{GeV}$, which serves as  a crude constraint to obey.

In the QCD sum rules for the hidden-charm tetraquark and pentaquark molecule  candidates, we usually choose the continuum threshold parameters as $\sqrt{s_0}=M_{gr}+(0.4\sim 0.7)\,\rm{GeV}$,  just like in the QCD sum rules for the traditional mesons, again the $gr$ denotes the ground states \cite{MNielsen-review-1812,WangZG-4-quark-mole,QCDSR-Pc-mole, WangZG-Landau}, though the hidden-charm molecule candidates  have not been unambiguously  assigned or determined   yet. In the present time, as the multiquark spectroscopy are poorly known,    we have to obtain predictions based on assumptions in one way or the other, then confront them to the experimental data in the future.

 After trial and error, we obtain the best continuum threshold parameters, Borel parameters, energy scales of the QCD spectral densities, and thereafter the pole contributions, which are shown explicitly in Table \ref{BCEPMR}. From the Table, we can see that the pole contributions are about $(40-60)\%$, the pole dominance criterion is well satisfied.

\begin{figure}
\centering
\includegraphics[totalheight=7cm,width=9cm]{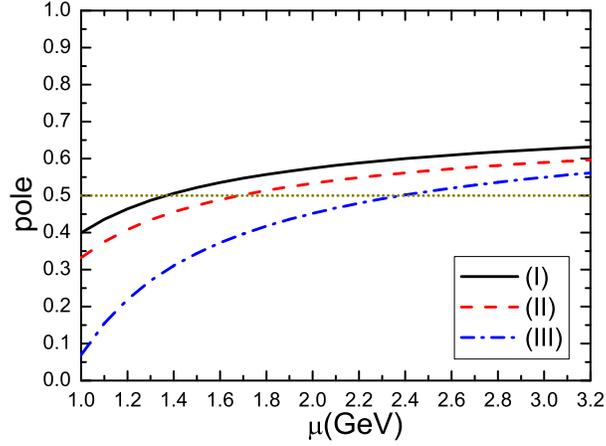}
  \caption{ The pole contributions with variations of the energy scales $\mu$ of the QCD spectral densities, where the (I), (II) and (III) correspond  to the currents
  $J_{c\bar{c}}(x)$, $J_{cc}(x)$ and $J_{\mu}^{cc\bar{c}}(x)$, respectively. The central values of the other parameters  are chosen.   }\label{Pole-mu}
\end{figure}

In Fig.\ref{Pole-mu}, we plot the pole contributions with variations of the energy scales $\mu$ of the QCD spectral densities for the currents
  $J_{c\bar{c}}(x)$, $J_{cc}(x)$ and $J_{\mu}^{cc\bar{c}}(x)$ with the central values of other input parameters shown in Table \ref{BCEPMR}. From the figure, we can see that the pole contributions increase monotonically and quickly with the increase of the energy scales of the QCD spectral densities before reaching  $50\%$, then they increase  monotonically but  slowly. The pole contributions exceed $50\%$ at the energy scales $\mu=1.5\,\rm{GeV}$, $1.8\,\rm{GeV}$ and $2.4\,\rm{GeV}$ for the currents $J_{c\bar{c}}(x)$, $J_{cc}(x)$ and $J_{\mu}^{cc\bar{c}}(x)$, respectively. The energy scale formula shown in Eq.\eqref{formula} plays a very  important role in enhancing the pole contributions.

\begin{figure}
\centering
\includegraphics[totalheight=6cm,width=7cm]{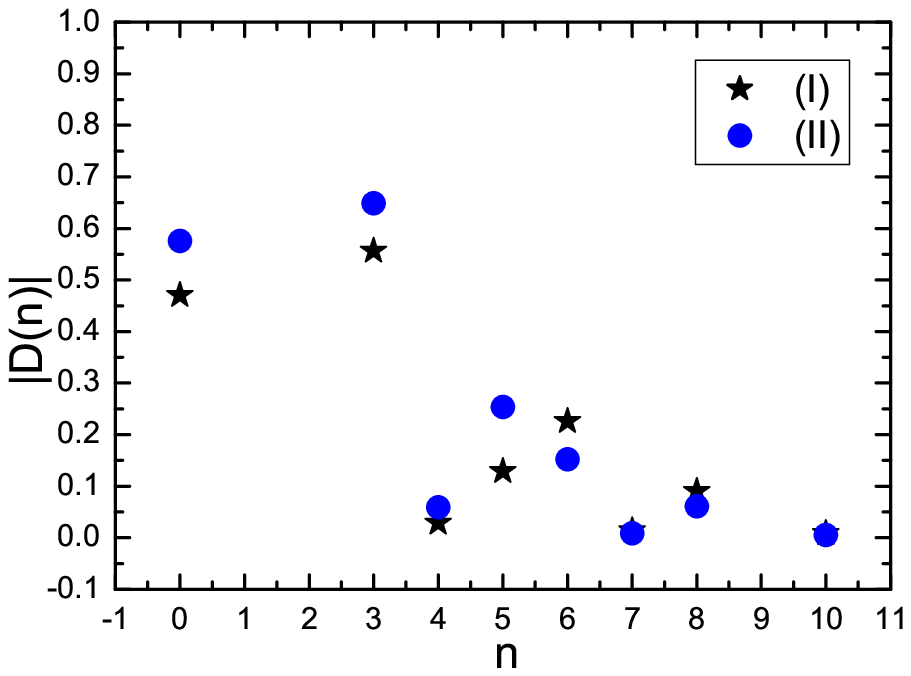}
\includegraphics[totalheight=6cm,width=7cm]{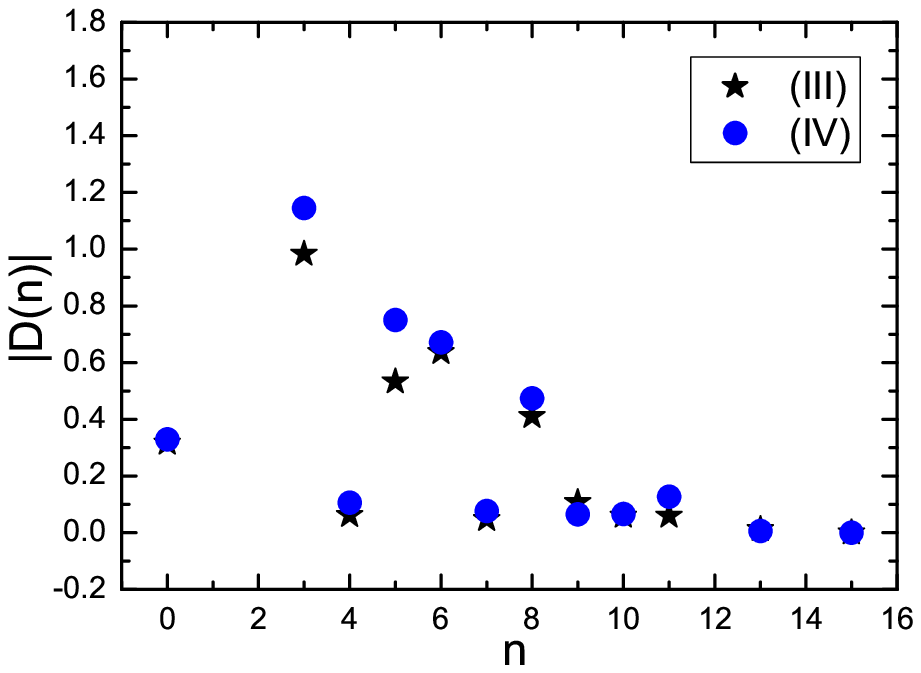}
  \caption{ The absolute values of the contributions  of the  vacuum condensates, where the (I), (II), (III)  and (IV) correspond  to the currents
  $J_{c\bar{c}}(x)$, $J_{cc}(x)$, $J_{\mu}^{cc\bar{c}}(x)$ and $J_{\mu}^{ccc}(x)$, respectively. The central values of the other parameters  are chosen.  }\label{OPE-cc-ccc}
\end{figure}

 In Fig.\ref{OPE-cc-ccc}, we plot the absolute values of the contributions  of the  vacuum condensates for the central values of the input parameters shown in Table \ref{BCEPMR} under the condition that the total contributions are normalized to be 1. From the figure, we can see that although the perturbative terms cannot make the dominant contributions, the operator product expansions  have very good  convergent behaviors.  The largest contributions come from the vacuum condensates $\langle\bar{q}q\rangle$, which  serve as  a milestone, the contributions of the vacuum condensates $D(n)$ decrease quickly with  the increase of the dimension $n$ except for some vibrations due to the tiny contributions  $D(4)$, $D(7)$ and large contributions $D(6)$. The contributions of the vacuum condensates $ \langle\frac{\alpha_sGG}{\pi}\rangle$ and $\langle\bar{q}q\rangle\langle\frac{\alpha_sGG}{\pi}\rangle$, which are of dimension $4$ and $7$ respectively, play a minor important role. For the currents $J_{c\bar{c}}(x)$ and $J_{\mu}^{cc\bar{c}}(x)$, $|D(6)|>|D(5)|$.

\begin{figure}
\centering
\includegraphics[totalheight=6cm,width=7cm]{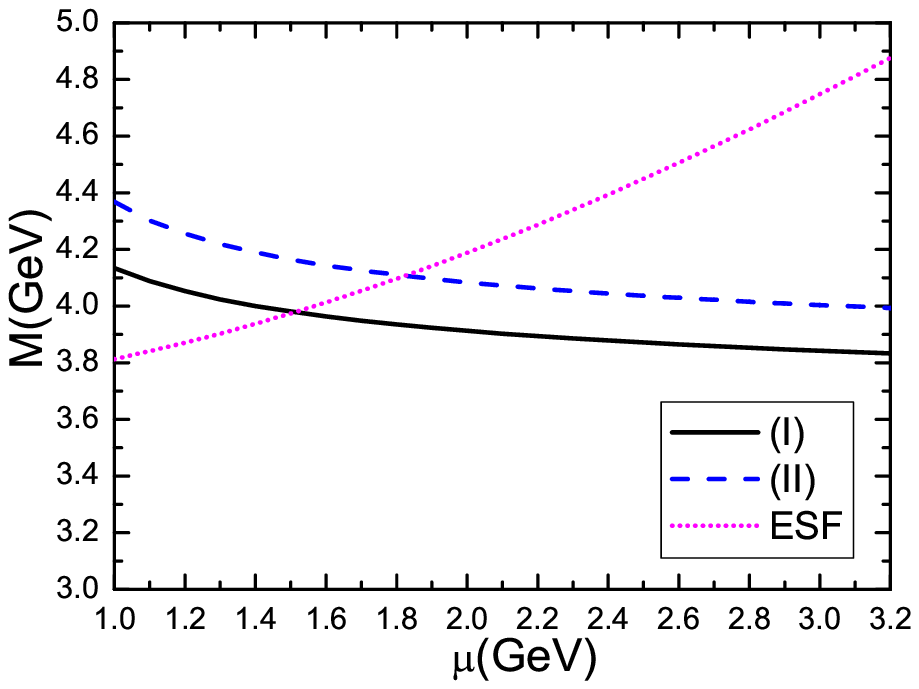}
\includegraphics[totalheight=6cm,width=7cm]{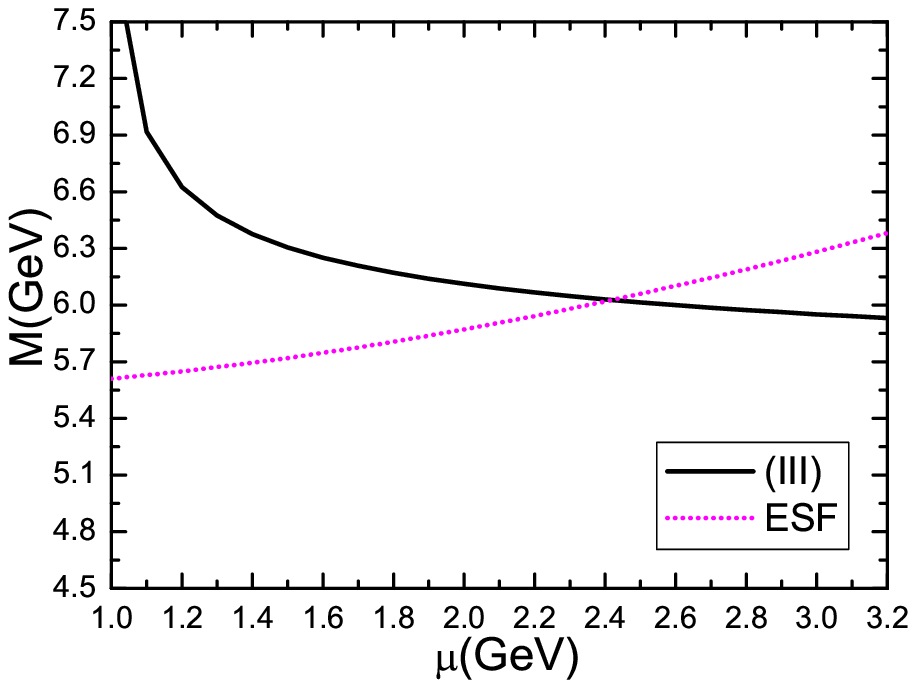}
  \caption{ The masses of the tetraquark and hexaquark molecular states with variations of the energy scales $\mu$ of the QCD spectral densities, where the (I), (II) and (III) correspond  to the  $D^*\bar{D}^*$, $D^*D^*$ and $D^*D^*\bar{D}^*$ tetraquark and hexaquark molecular  states, respectively,   the ESF denotes the formulas   $M=\sqrt{\mu^2+4\times (1.84\,\rm{GeV})^2}$ and $\sqrt{\mu^2+9\times (1.84\,\rm{GeV})^2}$, respectively.   }\label{mass-mu}
\end{figure}

In Fig.\ref{mass-mu}, we plot the predicted masses of the  $D^*\bar{D}^*$, $D^*D^*$ and $D^*D^*\bar{D}^*$ tetraquark and hexaquark molecular states
 with variations of the energy scales $\mu$ of the QCD spectral densities, where we have taken the central values of the input parameters.
On the other hand, we can rewrite the energy scale formulas as
\begin{eqnarray}
M_{T}&=&\sqrt{\mu^2+4{\mathbb{M}}_c^2}\, ,\nonumber\\
M_{H}&=& \sqrt{\mu^2+9{\mathbb{M}}_c^2}\, .
\end{eqnarray}
If we set ${\mathbb{M}}_c=m_c=1.84\,\rm{GeV}$, we can obtain the dash-dotted lines $M_{T}=\sqrt{\mu^2+4\times(1.84\,\rm{GeV})^2}$ and $M_{H}=\sqrt{\mu^2+9\times(1.84\,\rm{GeV})^2}$ in Fig.\ref{mass-mu}, which intersect with the lines of the masses of the
$D^*\bar{D}^*$, $D^*D^*$ and $D^*D^*\bar{D}^*$ tetraquark or hexaquark molecular states at the energy scales about $\mu=1.5\,\rm{GeV}$, $1.8\,\rm{GeV}$ and $2.4\,\rm{GeV}$, respectively. In this way, we choose the energy scales of the QCD spectral densities in a consistent way.

\begin{figure}
\centering
\includegraphics[totalheight=6cm,width=7cm]{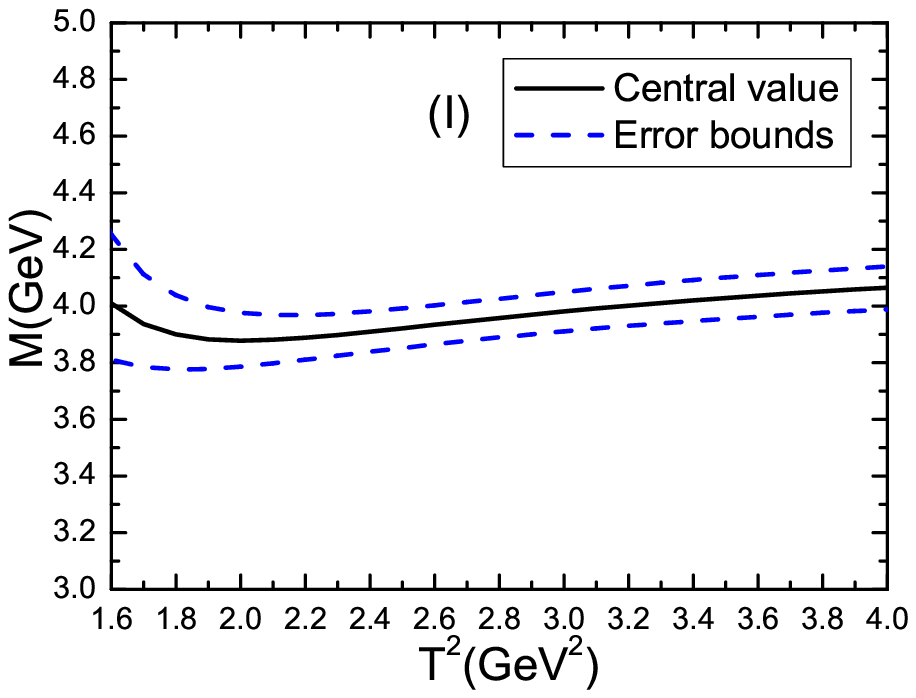}
\includegraphics[totalheight=6cm,width=7cm]{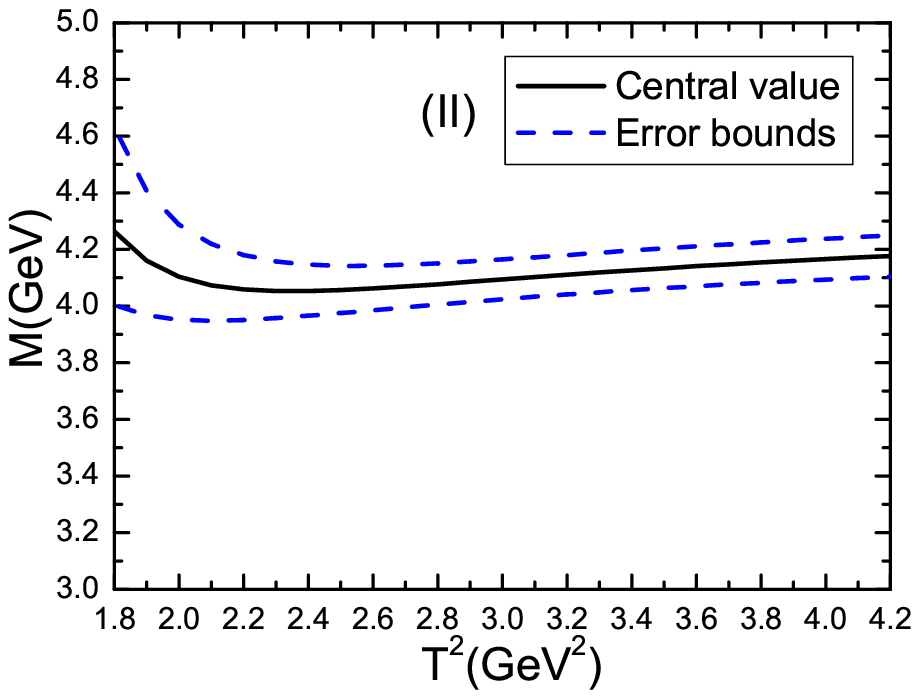}
\includegraphics[totalheight=6cm,width=7cm]{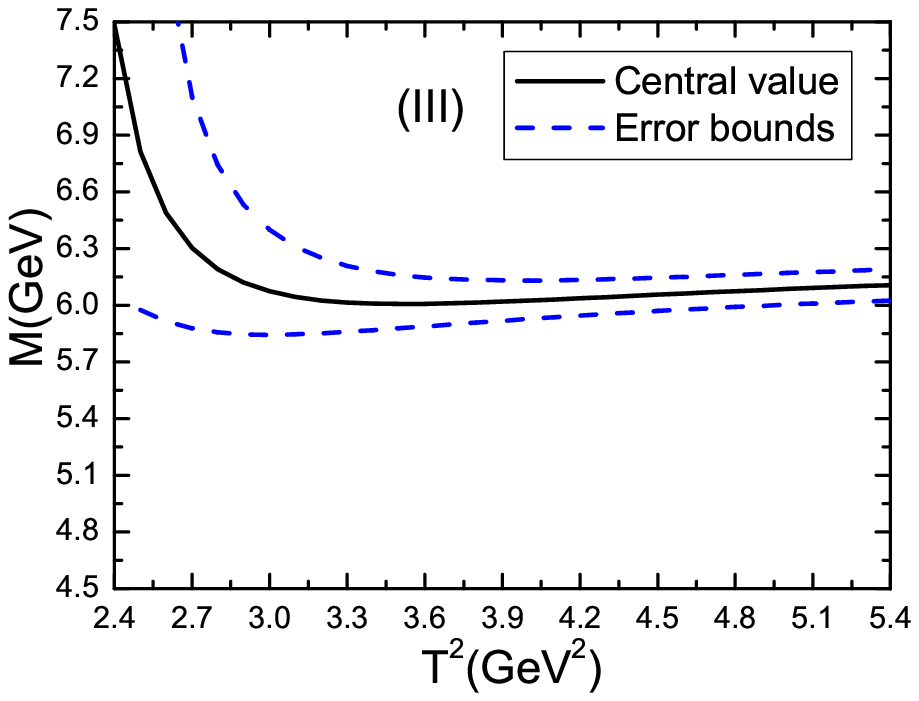}
\includegraphics[totalheight=6cm,width=7cm]{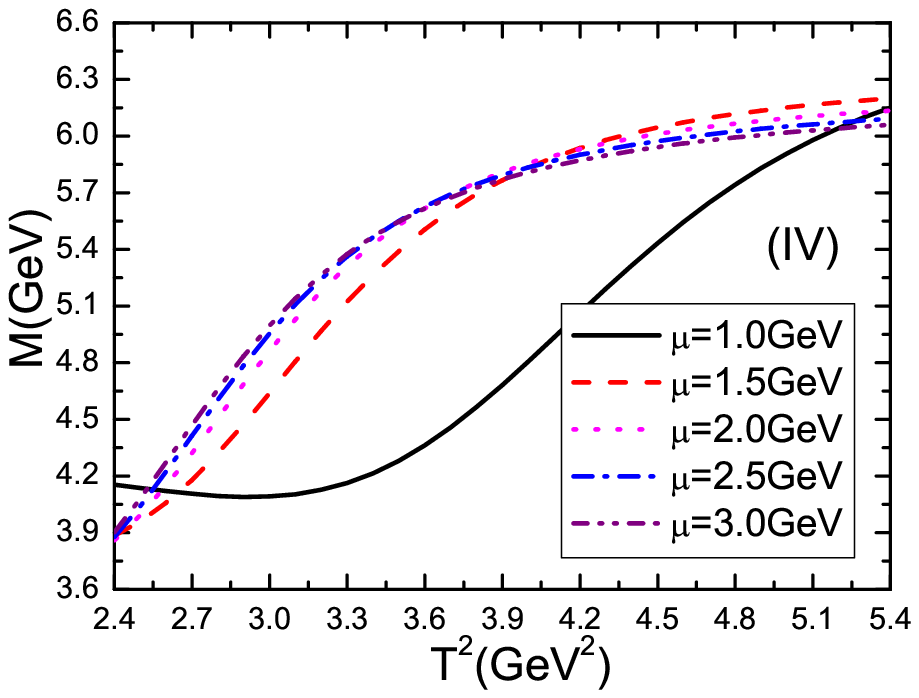}
  \caption{ The masses of the tetraquark and hexaquark molecular states with variations of the Borel parameters $T^2$, where the (I), (II), (III)  and (IV) correspond  to the $D^*\bar{D}^*$, $D^*D^*$, $D^*D^*\bar{D}^*$ and $D^*D^*D^*$ tetraquark and hexaquark molecular  states, respectively.    }\label{mass-cc-ccc}
\end{figure}

In this article, we choose the  value of the effective $c$-quark mass  ${\mathbb{M}}_c=1.84\pm 0.01\,\rm{GeV}$ \cite{WangZG-Landau},  which leads to a uncertainty  $\delta\mu=\pm0.1\,\rm{GeV}$ for the QCD spectral densities.  Now we take  into account all uncertainties of the input parameters,
and obtain the values of the masses and pole residues of
 the   $D^*\bar{D}^*$,  $D^*D^*$  and $D^*D^*\bar{D}^*$ tetraquark and hexaquark molecular states, which are  shown explicitly in Table \ref{BCEPMR-mass} and Fig.\ref{mass-cc-ccc}.

 From Fig.\ref{mass-cc-ccc}, we can see that there appear flat platforms  in the Borel windows   for the $D^*\bar{D}^*$,  $D^*D^*$  and $D^*D^*\bar{D}^*$ molecular states, it is reliable to extract the tetraquark and hexaquark molecular state masses.  We  can search for the scalar $D^*\bar{D}^*$,  $D^*D^*$ tetraquark states and the vector $D^*D^*\bar{D}^*$  hexaquark molecular state at the LHCb, Belle II,  CEPC, FCC, ILC  in the future.

We can take into account the  contributions of the intermediate   two-meson and three-meson  scattering states  to the correlation functions $\Pi_{T}(p^2)$ and $\Pi_{H}(p^2)$  according to the arguments presented  in Refs.\cite{WangZG-Landau,WangHuangtao-PRD,WangZG-IJMPA-4200}, as the currents $J_{c\bar{c}}(x)$, $ J_{cc}(x)$,
$J^{cc\bar{c}}_\mu(x)$ and  $J^{ccc}_\mu(x)$ also couple potentially to the $D^*\bar{D}^*$, $D^*D^*$, $D^*D^*\bar{D}^*$ and $D^*D^*D^*$ scattering states respectively according to the standard definition,
\begin{eqnarray}
\langle 0|\bar{q}(0)\gamma_{\alpha}c(0)|D^*(p)\rangle&=&f_{D^*}m_{D^*}\varepsilon_\alpha\, ,
\end{eqnarray}
 where the $\varepsilon_\alpha$ is the polarization vector of the $D^*$ meson.
The renormalized self-energies due to the intermediate meson-loops  contribute  a finite imaginary part to modify the dispersion relation,
\begin{eqnarray}
\Pi_{T/H}(p^2) &=&-\frac{\lambda_{T/H}^{2}}{ p^2-M_{T/H}^2+i\sqrt{p^2}\Gamma_{T/H}(p^2)}+\cdots \, .
 \end{eqnarray}
We take into account the finite width effects by the following simple replacement of the hadronic spectral densities,
\begin{eqnarray}
\lambda^2_{T/H}\delta \left(s-M^2_{T/H} \right) &\to& \lambda^2_{T/H}\frac{1}{\pi}\frac{M_{T/H}\Gamma_{T/H}(s)}{(s-M_{T/H}^2)^2+M_{T/H}^2\Gamma_{T/H}^2(s)}\, ,
\end{eqnarray}
then the hadron  sides of  the QCD sum rules in Eqs.\eqref{QCDSR-A}-\eqref{QCDSR-A-Deri} undergo the following changes,
\begin{eqnarray}
\lambda^2_{T/H}\exp \left(-\frac{M^2_{Z_c}}{T^2} \right) &\to& \lambda^2_{T/H}\int_{\Delta^2}^{s_0}ds\frac{1}{\pi}\frac{M_{T/H}\Gamma_{T/H}(s)}{(s-M_{T/H}^2)^2+M_{T/H}^2\Gamma_{T/H}^2(s)}\exp \left(-\frac{s}{T^2} \right)\, , \nonumber\\
&=&\tilde{\lambda}^2_{T/H}\exp \left(-\frac{M^2_{T/H}}{T^2} \right)\, , \\
\lambda^2_{T/H}M^2_{T/H}\exp \left(-\frac{M^2_{T/H}}{T^2} \right) &\to& \lambda^2_{T/H}\int_{\Delta^2}^{s_0}ds\,s\,\frac{1}{\pi}\frac{M_{T/H}\Gamma_{T/H}(s)}{(s-M_{T/H}^2)^2+M_{T/H}^2\Gamma_{T/H}^2(s)}\exp \left(-\frac{s}{T^2} \right)\, , \nonumber\\
&=& \tilde{\lambda}^2_{T/H}M^2_{T/H}\exp \left(-\frac{M^2_{T/H}}{T^2} \right)\, ,
\end{eqnarray}
where the $\Delta^2$ are  the two-meson or three-meson thresholds.  The  net effects  of the intermediate  meson-loops can be absorbed into the pole residues $\tilde{\lambda}_{T/H}$ safely without affecting the predicted tetraquark and hexaquark molecule masses. Even for the $Z_c(4200)$, the width is as large as $ 370^{+70}_{-70}{}^{+70}_{-132}\,\rm{MeV}$, the finite width effects can be safely absorbed into the pole residues \cite{WangZG-IJMPA-4200}, so
 the zero width approximation in  the hadronic spectral density  works very well.

In Fig.\ref{mass-cc-ccc}, we also plot the predicted masses  of the $D^*D^*D^*$ hexaquark molecular state with variations of the Borel parameters $T^2$ at  the energy scales of the QCD spectral densities $\mu=1.0\,\rm{GeV}$, $1.5\,\rm{GeV}$, $2.0\,\rm{GeV}$, $2.5\,\rm{GeV}$ and $3.0\,\rm{GeV}$. From the figure, we can see that there appear no platforms for the $D^*D^*D^*$ hexaquark molecular state. The QCD sum rules do not support the existence of the $D^*D^*D^*$ hexaquark molecular state with the $J^P=1^-$.

From Figs.\ref{DvDv-tree}-\ref{DvDv-cc-tree}, we can see that compared to the current $J_{c\bar{c}}(x)$, there are connected (nonfactorizable) Feynman diagrams in the correlation function $\Pi(p)$ for the current $J_{cc}(x)$ besides the disconnected (factorizable) Feynman diagrams. The connected diagrams and other diagrams obtained by substituting the quark lines with other terms in the full light and heavy propagators lead to more stable QCD sum rules for the predicted mass of the $D^*D^*$ tetraquark molecular state, see Fig.\ref{mass-cc-ccc}.
From Figs.\ref{DvDvDv-tree}-\ref{DvDvDv-ccc-tree-nonfact}, we can see that compared to the current $J_{cc\bar{c}}(x)$, there are connected Feynman diagrams in the correlation function $\Pi_{\mu\nu}(p)$ for the current $J_{ccc}(x)$ besides the disconnected Feynman diagrams. The contributions from the  connected diagrams and other diagrams obtained by substituting the quark lines with other terms in the full light and heavy propagators are so large as to make the  QCD sum rules for the mass of the $D^*D^*D^*$ hexaquark molecular state unstable, see Fig.\ref{mass-cc-ccc}.

In summary, the connected Feynman diagrams at the tree level shown in Fig.\ref{DvDv-cc-tree}  and their induced diagrams via substituting the quark lines
 make positive contributions,  the connected Feynman diagrams at the tree level shown  in Fig.\ref{DvDvDv-ccc-tree-nonfact} and their induced diagrams via substituting the quark lines  make negative or destructive  contributions, which are in contrast to  the assertion of  Lucha, Melikhov and Sazdjian that the factorizable (disconnected) diagrams in the color space only make contributions  to the  meson-meson scattering states, the tetraquark molecular states begin to receive contributions from the nonfactorizable (connected) diagrams at the order $\mathcal{O}(\alpha_s^2)$  \cite{Chu-Sheng-PRD-1,Chu-Sheng-PRD-2}. In fact, it is of no use to distinguish the factorizable and nonfactorizable properties of the Feynman diagrams in the color space.

\begin{figure}
\centering
\includegraphics[totalheight=6cm,width=7cm]{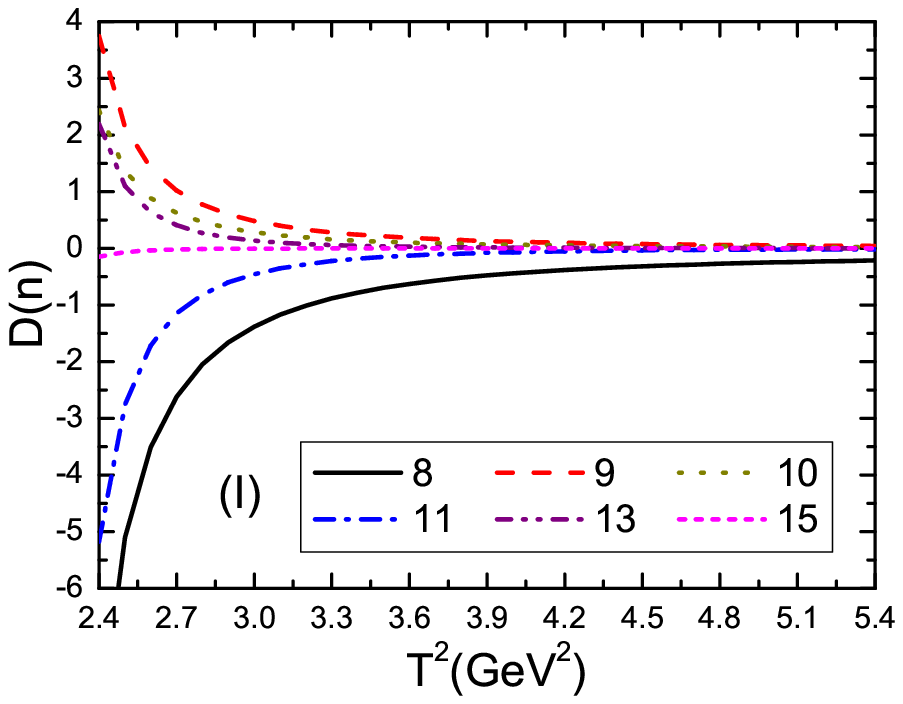}
\includegraphics[totalheight=6cm,width=7cm]{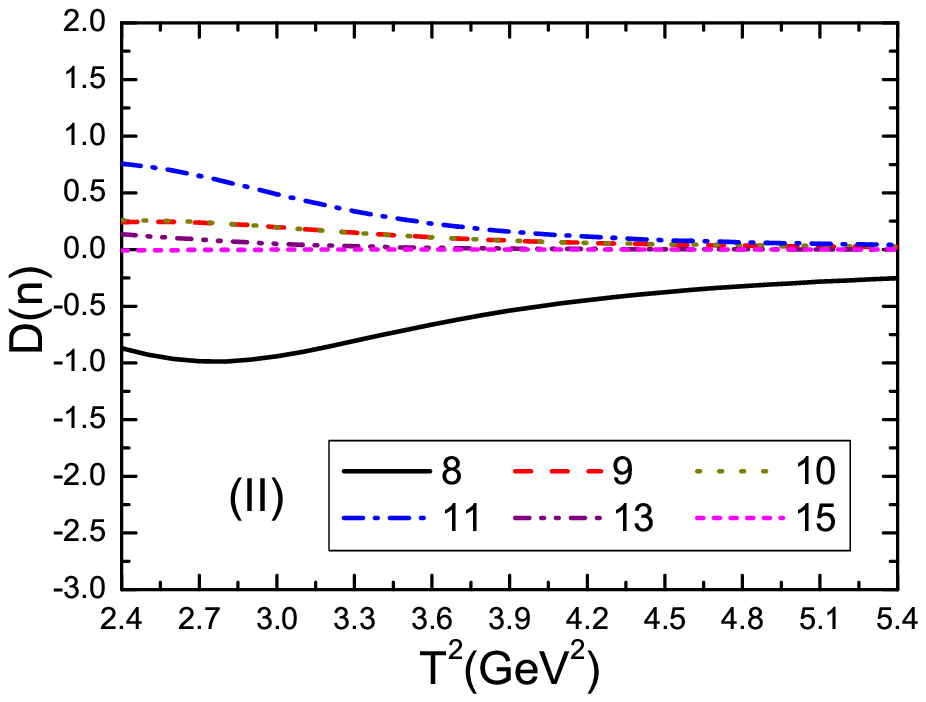}
  \caption{ The contributions of the higher dimensional vacuum condensates  with variations of the  Borel parameters $T^2$ for the central values of other parameters, where the (I) and (II) correspond  to the currents
   $J_{\mu}^{cc\bar{c}}(x)$ and $J_{\mu}^{ccc}(x)$, respectively.   }\label{OPE-cc-ccc-high}
\end{figure}

In Fig.\ref{OPE-cc-ccc-high}, we plot the contributions of the vacuum condensates  $D(n)$ with $n\geq 8$ for the $J_{\mu}^{cc\bar{c}}(x)$ and $J_{\mu}^{ccc}(x)$ with the central values of the input parameters shown in Table \ref{BCEPMR}.  From the figure, we can see that, for the current $J_{\mu}^{cc\bar{c}}(x)$, the contributions of the higher dimensional vacuum condensates are huge for the small Borel parameters $T^2$, and decrease monotonously and quickly  with the
 increase of the Borel parameters $T^2$ at the region $T^2\leq 3.6\,\rm{GeV}^2$, then  they decrease monotonously and slowly with the increase of the Borel parameters $T^2$.   The higher dimensional vacuum condensates play a minor important role in the Borel windows, but they play a very important role in determining   the Borel windows or in warranting the appearances of the Borel platforms. While for the current $J_{\mu}^{ccc}(x)$, the contributions of the higher dimensional vacuum condensates are not greatly amplified  for the small Borel parameters $T^2$, and  decrease monotonously and slowly with the increase of the Borel parameters $T^2$, which cannot stabilize the QCD sum rules to obtain reliable predictions, the QCD sum rules at the hadron side may be dominated by the three-meson scattering states.

\begin{table}
\begin{center}
\begin{tabular}{|c|c|c|c|c|c|c|c|}\hline\hline
$J^P$                       &$T^2(\rm{GeV}^2)$   &$\sqrt{s_0}(\rm{GeV})$   &$\mu(\rm{GeV})$  &pole           \\ \hline

$0^+(D^*\bar{D}^*)$         &$2.8-3.2$           &$4.55\pm0.10$            &$1.5$            &$(41-64)\%$    \\ \hline

$0^+(D^*D^*)$               &$3.0-3.4$           &$4.65\pm0.10$            &$1.8$            &$(41-62)\%$     \\ \hline

$1^-(D^*D^*\bar{D}^*)$      &$3.9-4.3$           &$6.60\pm0.10$            &$2.4$            &$(41-60)\%$     \\ \hline

$1^-(D^*D^*D^*)$            &$3.9-4.3$           &$6.60\pm0.10$            &$2.4$            &$(39-60)\%$    \\ \hline\hline
\end{tabular}
\end{center}
\caption{ The Borel parameters, continuum threshold parameters,   energy scales of the QCD spectral densities and pole contributions for the
 $D^*\bar{D}^*$, $D^*D^*$, $D^*D^*\bar{D}^*$ and $D^*D^*D^*$ tetraquark and hexaquark molecular  states. }\label{BCEPMR}
\end{table}

\begin{table}
\begin{center}
\begin{tabular}{|c|c|c|c|c|c|c|c|}\hline\hline
$J^P$                          &$M_{T/H}$                 &$\lambda_{T/H}$ \\ \hline

$0^+(D^*\bar{D}^*)$            &$3.98\pm0.09\,\rm{GeV}$   &$(4.05\pm0.70)\times 10^{-2}\rm{GeV}^5$  \\ \hline

$0^+(D^*D^*)$                  &$4.11\pm0.09\,\rm{GeV}$   &$(8.36\pm1.32)\times 10^{-2}\rm{GeV}^5$  \\ \hline

$1^-(D^*D^*\bar{D}^*)$         &$6.03\pm0.11\,\rm{GeV}$   &$(3.14\pm0.55)\times 10^{-3}\rm{GeV}^8$  \\ \hline \hline
\end{tabular}
\end{center}
\caption{ The   masses and pole residues for the $D^*\bar{D}^*$, $D^*D^*$ and $D^*D^*\bar{D}^*$  tetraquark and hexaquark molecular  states. }\label{BCEPMR-mass}
\end{table}

\section{Conclusion}
In this article,  we construct the color-singlet-color-singlet type currents and the color-singlet-color-singlet-color-singlet type currents to interpolate the scalar $D^*\bar{D}^*$, $D^*D^*$ tetraquark molecular states and the vector  $D^*D^*\bar{D}^*$, $D^*D^*D^*$ hexaquark molecular states, respectively, and study their masses and pole residues with the QCD sum rules in details by carrying out the operator product expansion up to the vacuum condensates of dimension 10 and dimension 15, respectively. In calculations, we choose the pertinent  energy scales of the QCD spectral densities  with the  energy scale formula $\mu=\sqrt{M^2_{T}-(2{\mathbb{M}}_c)^2}$  and $\sqrt{M^2_{H}-(3{\mathbb{M}}_c)^2}$ for the tetraquark molecular states and hexaquark molecular states respectively in a consistent way, which can enhance the pole contributions remarkably and also improve the convergent behaviors of the operator product expansion remarkably.
 We obtain stable QCD sum rules for the masses and pole residues of the scalar $D^*\bar{D}^*$, $D^*D^*$ tetraquark molecular states and the vector $D^*D^*\bar{D}^*$ hexaquark molecular state, but cannot obtain stable QCD sum rules for the vector $D^*D^*D^*$ hexaquark molecular state. We  can search for the  $D^*\bar{D}^*$, $D^*D^*$ and  $D^*D^*\bar{D}^*$ tetraquark and hexaquark molecular states  at the LHCb, Belle II,  CEPC, FCC, ILC  in the future.
 In calculations, we observe that the connected Feynman diagrams at the tree level and their induced diagrams via substituting the quark lines make positive contributions in the QCD sum rules for the $D^*D^*$ tetraquark molecular state, but make negative or destructive  contributions in the QCD sum rules for the $D^*D^*D^*$ hexaquark molecular state,  where the tree level denotes the lowest order contributions shown in Figs.\ref{DvDv-tree}-\ref{DvDvDv-ccc-tree-nonfact}. Lucha, Melikhov and Sazdjian assert that those nonfactorizable Feynman diagrams shown in Fig.\ref{DvDv-cc-tree} and Fig.\ref{DvDvDv-ccc-tree-nonfact}  can be deformed into the box diagrams. It is unfeasible, those nonfactorizable Feynman diagrams can only be deformed into the box diagrams in the color space, not in the momentum space.
 It is of no use or meaningless to distinguish the factorizable and nonfactorizable properties of the Feynman diagrams in the color space in the operator product expansion so as to interpret  them in terms of the hadronic observables,  we can only  obtain information about the short-distance and long-distance contributions.

\section*{Appendix}
The explicit expressions of the QCD spectral densities $\rho^{c\bar{c}}_{T}(s,\mu)$  and $\rho^{cc}_{T}(s,\mu)$,
\begin{eqnarray}
\rho^{c\bar{c}}_T(s,\mu)&=&\frac{3}{512\pi^6}\int  dy dz \,yz\left(1-y-z\right)^2\left(s-\overline{m}_c^2\right)^3\left(3s-\overline{m}_c^2\right) \nonumber\\
 &&-\frac{3m_c\langle \bar{q}q\rangle}{16\pi^4}\int dy dz  \,y \left(1-y-z\right)\left(s-\overline{m}_c^2\right)\left(2s-\overline{m}_c^2\right) \nonumber\\
 &&+\frac{m_c\langle \bar{q}g_s\sigma Gq\rangle}{64\pi^4}\int dydz  \,y \left(7s-6\overline{m}_c^2\right)+\frac{m_c^{2}\langle \bar{q}q\rangle^{2}}{4\pi^2}\int dy \,  \nonumber\\
&&-\frac{m_c^{2}\langle \bar{q}q\rangle\langle \bar{q}g_s\sigma Gq\rangle}{8\pi^2}\int dy \,\left(1+\frac{s}{T^{2}}\right)\delta \left(s-\widetilde{m}_c^2\right) \nonumber\\
&&+\frac{m_c^{2}\langle \bar{q}g_s\sigma Gq\rangle^2}{64\pi^2 T^6}\int dy  \, s^{2} \delta \left(s-\widetilde{m}_c^2\right)\nonumber\\
&&-\frac{m_c^2}{128\pi^4}\langle\frac{\alpha_{s}GG}{\pi}\rangle\int dy dz \,\frac{z(1-y-z)^2}{y^{2}} \left(3s-2\overline{m}_c^2\right)\nonumber\\
&&+\frac{m_c^3\langle \bar{q}q\rangle}{96\pi^2}\langle\frac{\alpha_{s}GG}{\pi}\rangle \int dydz \,\left(1+\frac{z}{y}\right)\frac{(1-y-z)}{y^{2}}\left(1+\frac{s}{T^2} \right)\delta \left(s-\overline{m}_c^2\right)\nonumber\\
&&+\frac{m_c\langle \bar{q}q\rangle}{32\pi^2}\langle\frac{\alpha_{s}GG}{\pi}\rangle\int dydz \,\left[1-\frac{z(1-y-z)}{y^{2}}\right]\left[2+s\,\delta\left(s-\overline{m}_c^2\right) \right] \nonumber\\
&&-\frac{m_c^4\langle \bar{q}q\rangle^{2}}{36T^4}\langle\frac{\alpha_{s}GG}{\pi}\rangle\int dy \,\frac{1}{y^{3}}\delta \left(s-\widetilde{m}_c^2\right)+\frac{m_c^{2}\langle \bar{q}q\rangle^{2}}{72T^6}\int dy  \, s^{2} \delta \left(s-\widetilde{m}_c^2\right)\nonumber\\
&&+\frac{m_c^2\langle \bar{q}q\rangle^2}{12T^{2}}\langle\frac{\alpha_{s}GG}{\pi}\rangle\int dy\,\frac{1}{y^{2}}\delta \left(s-\widetilde{m}_c^2\right)+\frac{\langle \bar{q}g_s\sigma Gq\rangle^2}{256\pi^2T^{2}}\int dy  \,s\delta \left(s-\widetilde{m}_c^2\right)\nonumber\\
&&-\frac{m_c\langle \bar{q}q\rangle}{192\pi^2}\langle\frac{\alpha_{s}GG}{\pi}\rangle\int dy\,y\left[2+s\delta \left(s-\widetilde{m}_c^2\right)\right]\, ,
\end{eqnarray}

\begin{eqnarray}
\rho^{cc}_{T}(s,\mu)&=&\frac{7}{512\pi^6}\int dydz \,yz\left(1-y-z\right)^2\left(s-\overline{m}_c^2\right)^3\left(3s-\overline{m}_c^2\right) \nonumber\\
 &&-\frac{m_c^{2}}{512\pi^6}\int dydz  \left(1-y-z\right)^2\left(s-\overline{m}_c^2\right)^3 \nonumber\\
 &&-\frac{m_c\langle \bar{q}q\rangle}{2\pi^4}\int dydz  \,y\left(1-y-z\right)\left(s-\overline{m}_c^2\right)\left(2s-\overline{m}_c^2\right)  \nonumber\\
&&+\frac{m_c\langle \bar{q}g_s\sigma Gq\rangle}{8\pi^4}\int dydz  \,y\left(3s-2\overline{m}_c^2\right) \nonumber\\
&&-\frac{\langle \bar{q}q\rangle^2}{24\pi^2}\int dy  \,y(1-y) \left(3s-2\widetilde{m}_c^2\right)+\frac{7m_c^{2}\langle \bar{q}q\rangle^2}{12\pi^2}\int dy \, \nonumber\\
&&+\frac{\langle \bar{q}q\rangle\langle \bar{q}g_s\sigma Gq\rangle}{48\pi^2}\int dy \,y(1-y)\left[6+\left(4s+\frac{s^2}{T^{2}}\right)\delta \left(s-\widetilde{m}_c^2\right)\right] \nonumber\\
&&-\frac{7m_c^{2}\langle \bar{q}q\rangle\langle \bar{q}g_s\sigma Gq\rangle}{24\pi^2}\int dy \,\left(1+\frac{s}{T^{2}}\right)\delta \left(s-\widetilde{m}_c^2\right)\nonumber\\
&&-\frac{\langle \bar{q}g_s\sigma Gq\rangle^2}{384\pi^2}\int dy  \,y(1-y)\left(6+\frac{6s}{T^2}+\frac{3s^{2}}{T^4}+\frac{s^{3}}{T^6} \right)\delta (s-\widetilde{m}_c^2)\nonumber\\
&&+\frac{7m_c^2\langle \bar{q}g_s\sigma Gq\rangle^2}{192\pi^2T^6}\int dy  \, s^{2} \delta \left(s-\widetilde{m}_c^2\right)\nonumber\\
&&-\frac{7m_c^2}{384\pi^4}\langle\frac{\alpha_{s}GG}{\pi}\rangle\int dydz \,\frac{z(1-y-z)^2}{y^{2}}\left(3s-2\overline{m}_c^2\right)\nonumber\\
&&+\frac{m_c^4}{768\pi^4}\langle\frac{\alpha_{s}GG}{\pi}\rangle\int  dydz \,\frac{(1-y-z)^2}{y^{3}}\nonumber\\
&&+\frac{m_c^3\langle \bar{q}q\rangle}{36\pi^2}\langle\frac{\alpha_{s}GG}{\pi}\rangle\int dydz \,\left(1+\frac{z}{y}\right)\frac{1-y-z}{y^{2}}\left(1+\frac{s}{T^2} \right)\delta (s-\overline{m}_c^2)\nonumber\\
&&+\frac{m_c^2\langle \bar{q}q\rangle^2}{216T^4}\langle\frac{\alpha_{s}GG}{\pi}\rangle\int dy\,\frac{1-y}{y^{2}}s\, \delta \left(s-\widetilde{m}_c^2\right)\nonumber\\
&&-\frac{7m_c^4\langle \bar{q}q\rangle^2}{108T^4}\langle\frac{\alpha_{s}GG}{\pi}\rangle\int dy\,\frac{1}{y^{3}} \delta \left(s-\widetilde{m}_c^2\right)\nonumber\\
&&-\frac{m_c^2}{256\pi^4}\langle\frac{\alpha_{s}GG}{\pi}\rangle\int dydz \,\frac{(1-y-z)^2}{y^{2}} \left(s-\overline{m}_c^2\right)\nonumber\\
&&+\frac{m_c\langle \bar{q}q\rangle}{48\pi^2}\langle\frac{\alpha_{s}GG}{\pi}\rangle\int dydz  \,\left[4-\frac{4z(1-y-z)}{y^{2}}+\frac{1-y-z}{z}-\frac{z}{y}\right]\left[2+s\delta\left(s-\overline{m}_c^2\right)\right]\nonumber\\
&&+\frac{7m_c^{2}\langle \bar{q}q\rangle^2}{36T^2}\langle\frac{\alpha_{s}GG}{\pi}\rangle\int dy\,\frac{1}{y^{2}}\delta \left(s-\widetilde{m}_c^2\right)\nonumber\\
&&-\frac{1}{128\pi^4}\langle\frac{\alpha_{s}GG}{\pi}\rangle\int dydz  \,yz \left(s-\overline{m}_c^2\right) \left(2s-\overline{m}_c^2\right)  \nonumber\\
&&-\frac{m_c^2}{256\pi^4}\langle\frac{\alpha_{s}GG}{\pi}\rangle\int dydz  \,\frac{4-3y-4z}{y}\left(s-\overline{m}_c^2\right) \nonumber\\
&&-\frac{1}{256\pi^4}\langle\frac{\alpha_{s}GG}{\pi}\rangle\int dydz  \,(1-y-z)^{2}\left(s-\overline{m}_c^2\right)\left(2s-\overline{m}_c^2\right)\nonumber\\
&&-\frac{\langle \bar{q}q\rangle^{2}}{144}\langle\frac{\alpha_{s}GG}{\pi}\rangle\int dy \,\left(1+\frac{5s}{T^{2}}\right)\delta \left(s-\widetilde{m}_c^2\right) \nonumber\\
&&-\frac{m_c\langle \bar{q}g_s\sigma Gq\rangle}{32\pi^4}\int dydz  \,\left(1-2y-z\right) \left(3s-2\overline{m}_c^2\right)\nonumber
\end{eqnarray}
\begin{eqnarray}
&&-\frac{\langle \bar{q}q\rangle\langle \bar{q}g_s\sigma Gq\rangle}{24\pi^2}\int dy \,\left(1-y\right)\left[2+s\,\delta \left(s-\widetilde{m}_c^2\right)\right]\nonumber\\
&&+\frac{\langle \bar{q}g_s\sigma Gq\rangle^2}{96\pi^2}\int dy \,\left(1-y\right)\left(2+\frac{2s}{T^2}+\frac{s^{2}}{T^4} \right)\delta \left(s-\widetilde{m}_c^2\right) \nonumber\\
&&-\frac{\langle \bar{q}g_s\sigma Gq\rangle^2}{192\pi^2}\int dy \,\left(1+\frac{s}{T^2} \right)\delta \left(s-\widetilde{m}_c^2\right)+\frac{ \langle \bar{q}g_s\sigma Gq\rangle^2}{128\pi^2T^{2}}\int dy \,s\,\delta \left(s-\widetilde{m}_c^2\right)\nonumber\\
&&-\frac{m_c\langle \bar{q}q\rangle}{72\pi^2}\langle\frac{\alpha_{s}GG}{\pi}\rangle\int dy \,y\left[2+s\,\delta \left(s-\widetilde{m}_c^2\right)\right] \nonumber\\
&&-\frac{\langle \bar{q}q\rangle^{2}}{432}\langle\frac{\alpha_{s}GG}{\pi}\rangle\int dy \,y(1-y)\left(6+\frac{6s}{T^2}+\frac{3s^{2}}{T^4}+\frac{s^{3}}{T^6}\right)\delta \left(s-\widetilde{m}_c^2\right) \nonumber\\
&&+\frac{7m_c^{2}\langle \bar{q}q\rangle^{2}}{216T^6}\langle\frac{\alpha_{s}GG}{\pi}\rangle\int dy \,s^{2}\delta \left(s-\widetilde{m}_c^2\right)\, ,
\end{eqnarray}
where $\int dydz=\int_{y_i}^{y_f}dy\int_{z_i}^{1-y}dz$, $\int dy=\int_{y_i}^{y_f}dy$,
$y_{f}=\frac{1+\sqrt{1-4m_c^2/s}}{2}$,
$y_{i}=\frac{1-\sqrt{1-4m_c^2/s}}{2}$, $z_{i}=\frac{y
m_c^2}{y s -m_c^2}$, $\overline{m}_c^2=\frac{(y+z)m_c^2}{yz}$,
$ \widetilde{m}_c2=\frac{m_c^2}{y(1-y)}$, $\int_{y_i}^{y_f}dy \to \int_{0}^{1}dy$, $\int_{z_i}^{1-y}dz \to \int_{0}^{1-y}dz$ when the $\delta$ functions $\delta\left(s-\overline{m}_c^2\right)$ and $\delta\left(s-\widetilde{m}_c^2\right)$ appear.

\section*{Acknowledgements}
This  work is supported by National Natural Science Foundation, Grant Number  11775079.

\end{document}